\documentclass{article}
\usepackage{graphicx} 
\usepackage{authblk}
\usepackage[margin=1in]{geometry}
\usepackage{amsmath, amssymb}
\usepackage{dirtytalk}
\usepackage{hyperref}
\usepackage[english]{babel}
\usepackage[toc,page]{appendix}
\usepackage{subcaption}
\usepackage[x11names, svgnames, dvipsnames]{xcolor}
\usepackage{wrapfig}
\usepackage{dirtytalk}
\usepackage{lineno}
\usepackage[T1]{fontenc}
\usepackage{tikz}
\usepackage{cite}
\usepackage{float}
\usepackage{enumitem}
\usepackage{multirow}
\usepackage{tabularx}
\usepackage{array}
\usetikzlibrary{trees}
\usepackage{algorithm}
\usepackage{algpseudocode}
\newcommand{\keywords}[1]{\par\addvspace\baselineskip
\noindent\textbf{Keywords:}\enspace\ignorespaces#1}

\title{BASIN: Bayesian mAtrix variate normal model with Spatial and sparsIty priors in Non-negative deconvolution}

\author[1]{Jiasen Zhang}
\author[2]{Xi Qiao}
\author[2, *]{Liangliang Zhang}
\author[1, *]{Weihong Guo}

\affil[1]{Department of Mathematics, Applied Mathematics and Statistics, Case Western Reserve University, Cleveland, OH}

\affil[2]{Department of Population and Quantitative Health Sciences, Case Western Reserve University, Cleveland, OH}

\affil[*]{Corresponding author: lxz716@case.edu, weihong.guo@case.edu}

\date{}

\begin{document}

\maketitle

\abstract{Spatial transcriptomics allows researchers to visualize and analyze gene expression within the precise location of tissues or cells. It provides spatially resolved gene expression data but often lacks cellular resolution, necessitating cell type deconvolution to infer cellular composition at each spatial location. In this paper we propose BASIN for cell type deconvolution, which models deconvolution as a nonnegative matrix factorization (NMF) problem incorporating graph Laplacian prior. Rather than find a deterministic optima like other recent methods, we propose a matrix variate Bayesian NMF method with nonnegativity and sparsity priors, in which the variables are maintained in their matrix form to derive a more efficient matrix normal posterior. BASIN employs a Gibbs sampler to approximate the posterior distribution of cell type proportions and other parameters, offering a distribution of possible solutions, enhancing robustness and providing inherent uncertainty quantification. The performance of BASIN is evaluated on different spatial transcriptomics datasets and outperforms other deconvolution methods in terms of accuracy and efficiency. The results also show the effect of the incorporated priors and reflect a truncated matrix normal distribution as we expect.}



\section{Introduction}\label{introduction}
Spatial transcriptomics (ST) is a cutting-edge technique in the field of genomics that allows researchers to analyze the gene expression profile of tissues while preserving spatial information \cite{tian2023expanding}. It offers insights into the spatial distribution of gene expression within individual cells or regions of a tissue, and it has many exciting applications in fields such as developmental biology, cancer research, neurobiology, nephrology, and immunology \cite{rao2021exploring, moses2022museum, jin2024advances, jain2024spatial, lein2017promise}. ST technology can be performed at either single-cell or spot-level resolution. Obviously, single-cell resolution ST provides higher granularity for downstream analyses and facilitates more accurate biological interpretations, but it is more expensive. Spot resolution ST technology, such as 10x Visium, is cheaper and more prevalent. Typically, 10x Visium data includes three primary types of information: gene expression data, spatial information encoded by barcodes, and histological images, which enables simultaneous integration of molecular and imaging data generated from Formalin-Fixed Paraffin Embedded (FFPE) or frozen tissue sections. However, a main limitation of spot resolution ST data is its inability to precisely map molecular profiles to individual cells, as each spot captures the transcriptomes of roughly 10–200 cells depending on the tissue context \cite{saiselet2020transcriptional}. To address this, inferring the cell-type composition at each spatial location typically requires a process known as cell type deconvolution \cite{zhang2020spatial} or spatial decomposition \cite{zeng2022statistical}. It involves employing computational and statistical models to decipher the intricate patterns of gene expression within tissues and infer specific cell types accurately.

To achieve single-cell level recognition within each spatial spot, researchers typically integrate paired droplet-based single-cell RNA-seq (scRNA-seq) profiles into their studies. It is crucial for unraveling the intricacies inherent in ST as scRNA-seq data provides a more detailed and comprehensive view of individual cell profiles and thus bridges the gap between spatial information and detailed cellular characterization. This enables the identification of cell types and facilitates the learning of the distribution of signals at a granular level. With different ways to represent and analyze ST data, recent deconvolution methods can be broadly grouped as follows: nonnegative matrix factorization (NMF)-based methods such as SPOTlight \cite{elosua2021spotlight} and NMFreg \cite{rodriques2019slide}, regression and Bayesian model–based methods such as CARD \cite{ma2022spatially}, SpatialDWLS \cite{dong2021spatialdwls}, MuSiC \cite{Wang2019}, RCTD \cite{cable2022robust}, Cell2location \cite{kleshchevnikov2022cell2location}, DestVI \cite{lopez2022destvi}, SpatialDecon \cite{danaher2022advances}, Stereoscope \cite{andersson2020single} and STRIDE \cite{sun2022stride}, optimal transport (OT)-based methods such as SpaOTsc \cite{cang2020inferring} and novoSpaRc \cite{moriel2021novosparc}, graph convolutional network (GCN)-based methods such as DSTG \cite{song2021dstg} and SD$^2$ \cite{li2022sd2}, and deep learning-based methods such as TANGRAM \cite{biancalani2021deep} and SpaDecon \cite{coleman2023spadecon}. There are also reference-free methods that do not require scRNA-seq data as a reference, but try to simultaneously find the cell type composition and gene expression profiles of each cell type \cite{ma2022spatially, miller2022reference, andrade2021bayesian, berglund2018spatial, zhang2025flexible, hu2021spagcn, zhao2021spatial}. Although there is no need for external data, reference-free methods are naturally less accurate and can produce ambiguous or biologically uninterpretable components.

In this paper, we propose BASIN for spatial transcriptomics cell type deconvolution using scRNA-seq data as a reference. Our contributions and novelties can be summarized as follows:
\begin{enumerate}
\item First, we represent cell type deconvolution in NMF model and incorporate graph Laplacian regularity utilizing both spatial information and histology images that are often collected together with ST data. It has been shown that in spatial transcriptomics, similar cell types have been shown to co-localize, with the co-localization pattern decaying as distance increases \cite{ma2022spatially}. However, most of the methods mentioned above did not consider this important spatial information. In addition, there is also no thorough study on how to better use the histological intensity images that come with ST for better performance. High-quality  histology images clearly show different regions of the tissue and provide informative clustering guidance. 

\item Second, unlike other NMF-based and probabilistic model-based methods that find deterministic solutions based on methods such as MAP estimation and gradient descent, we adopt a Bayesian framework and derive a Gibbs sampler to approximate the posterior distribution of the parameters and variables involved. As a result, BASIN adds an interpretable randomness to the model and helps enhance robustness to local optima and enable quantification of uncertainty. 

\item Third, none of the methods mentioned above incorporates sparsity in their models. As a result, they may return too uniformly distributed proportions and fail to distinguish different regions on the tissue. As a solution, we exploit Laplacian distribution as a prior to induce sparsity in our Bayesian model. When replaced with an exponential prior, it retains the sparsity-inducing property while inherently enforcing non-negativity.

\item In computational aspects, we propose a matrix variate Bayesian NMF method to derive the posteriors of the parameters and variables. In previous Bayesian NMF methods \cite{lu2022flexible, schmidt2009bayesian}, the matrices are vectorized to derive a multivariate normal posterior distribution element-wisely. However, we keep all the variables in their matrix form and derive a matrix normal posterior distribution, which enables the combination of Bayesian model and graph Laplacian prior, and improve sampling efficiency.
\end{enumerate}

We evaluate the performance of BASIN on different simulation studies and several published real spatial transcriptomics datasets, and compared it with some other recently proposed methods with high citations. We also illustrate the correctness of the derived posterior distributions and how it benefits the results.

\section{Results}\label{results}
\subsection{Simulation study}

\begin{figure}[h]
\centering
\includegraphics[width=0.9\textwidth]{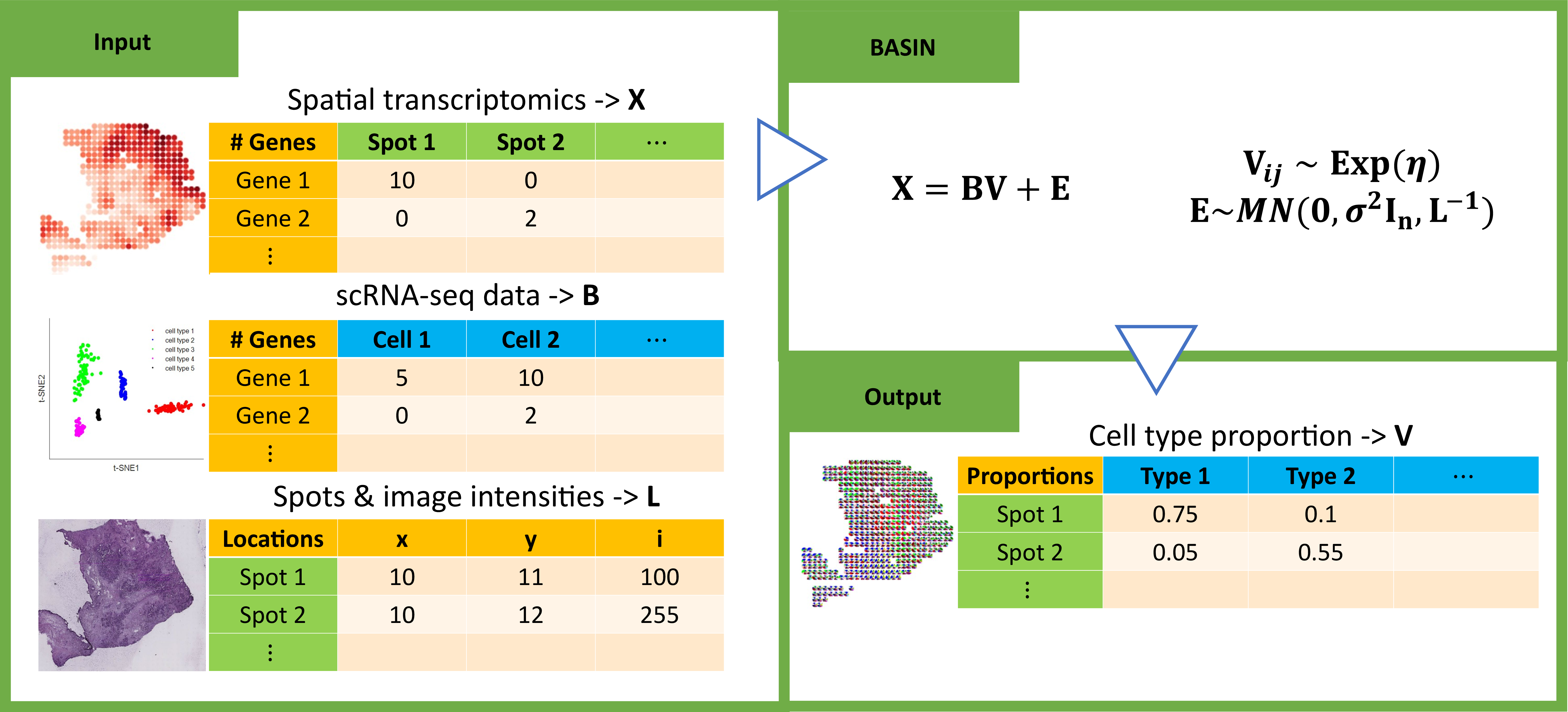}
\caption{Workflow of BASIN. BASIN starts from a spatial transcriptomics data (the first row), which is the target data to be deconvolved, and a scRNA-seq data (the third row), which is a reference to extract cell type information. To utilize the spatial structure in the ST data, we calculate the Graph Laplacian based on the coordinate information and the histology image (the second row). Besides, the mean gene expression for each cell type is computed from the scRNA-seq data as another input, which largely improve the computational efficiency and reduce memory usage. The outputs (right) are sampled from the posterior distribution, indicating the cell type proportions of each location.}\label{overview}
\end{figure}

The workflow of BASIN is summarized in Fig.~\ref{overview}. The details of BASIN, and data processing are discussed in Methods. To evaluate the performance and capability of BASIN, we perform four numerical experiments using simulated data and compare our results with five deconvolution methods: Stereoscope, SPOTlight, SpatialDWLS, RCTD and CARD. In summary, the simulation is based on spatial transcriptomics and scRNA sequence data from a pancreatic ductal adenocarcinoma tumor sample (PDAC-A) \cite{moncada2020integrating}, with three dominant cell types: acinar cells, cancer cluster 1, and terminal ductal cells, each predominantly localized in a distinct region. In the different cases, we assume that the cell type proportions are piecewise constant (Simulation 1), piecewise smooth (Simulation 2 and 3), and Dirichlet distributed (Simulation 4). We evaluated the results in three metrics: root mean squared error (RMSE), structural similarity index measure (SSIM), and Jensen-Shannon distance (JSD). The details of the simulation strategy, the compared methods, and the evaluation metrics are described in Methods.
\begin{figure}[h]
\centering
\includegraphics[width=0.9\textwidth]{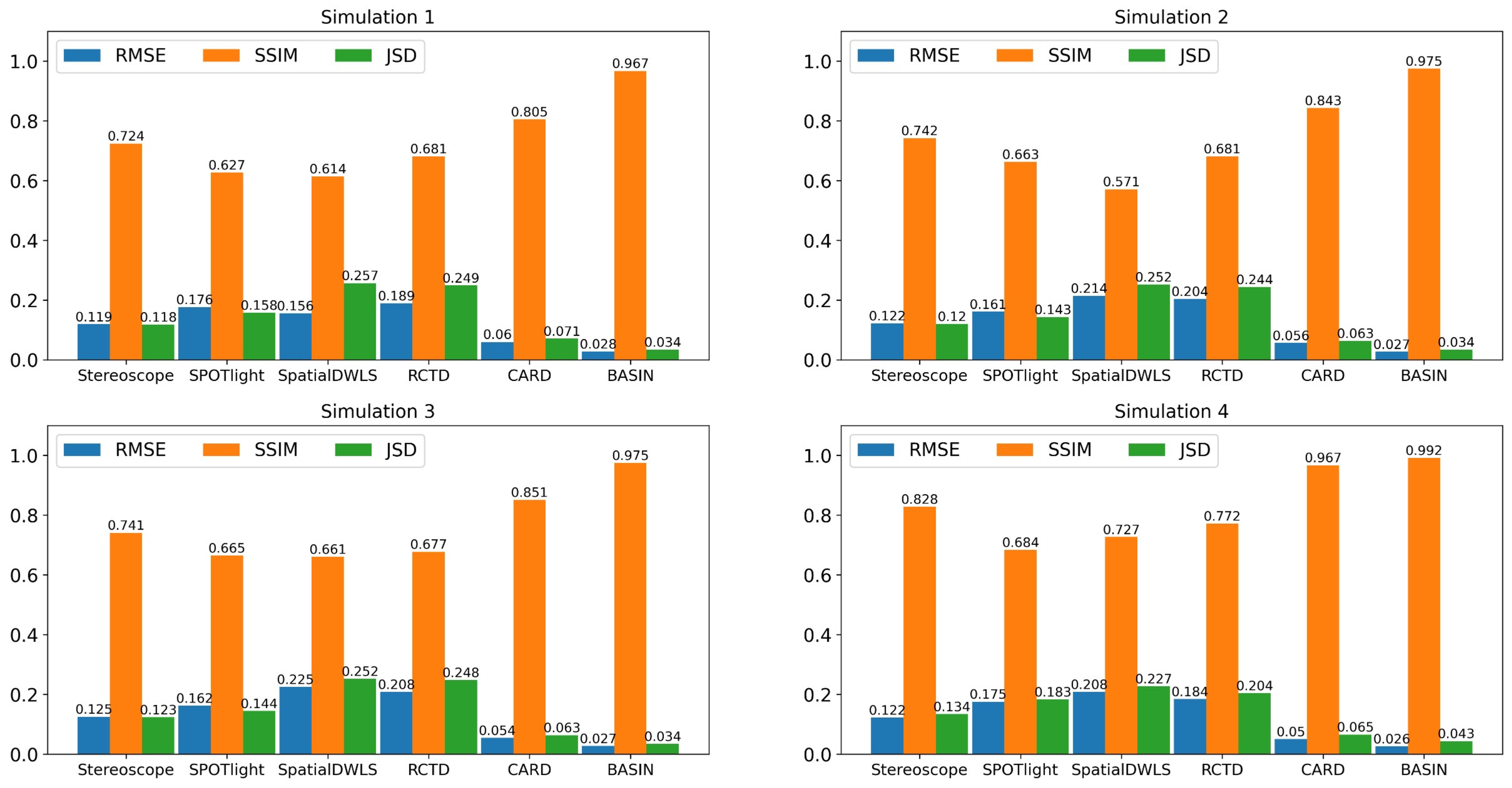}
\caption{Performance of different methods on the simulated data. Simulation 1: Each region has a piecewise constant cell type composition. Simulation 2: The simulated cell type proportion data is smoothed based on simulation 1. Simulation 3: The cell type proportions follow Gaussian distributions smoothed with a Gaussian kernel. Simulation 4: The three cell types follow different Dirichlet distributions smoothed by a Gaussian kernel. RMSE: Root mean square error, the the lower the better. SSIM: Structural similarity index measure, the higher the better. JSD: Jensen-Shannon distance, the lower the better.}\label{sim}
\end{figure}

The results shown in Fig.~\ref{sim} indicate that the proposed method consistently outperforms other methods in all simulations and metrics. In simulation 1, we set the cell type proportions to be piecewise constant to accurately evaluate the fitting ability. In this simple case, the proportions of all the locations consist of 0.7, 0.15 and 0.15 with clear but complicated boundaries among different regions (Supplementary Fig.~\ref{sim_regions}). BASIN achieves the lowest RMSE (0.028), compared with Stereoscope (0.119), SPOTlight (0.176), SpatialDWLS (0.156), RCTD (0.189) and CARD (0.06). BASIN also performs the best in terms of SSIM (0.967) compared with Stereoscope (0.724), SPOTlight (0.627), SpatialDWLS (0.614), RCTD (0.681) and CARD (0.805). Besides, BASIN has much lower JSD (0.034) than Stereoscope (0.118), SPOTlight (0.158), SpatialDWLS (0.257), RCTD (0.249) and CARD (0.071).

For the other three simulations, we add more conditions to simulate different properties that could happen in real data. In simulation 2, we simulate the spatial correlation and the cell type proportions are no longer piecewise constant but change smoothly. In simulation 3, we add noise to the cell type proportions based on simulation 2 to evaluate robustness and the ability of recovering more complicated spatial structure. In simulation 4, we implement a new simulation assuming that the cell type proportions follow Dirichlet distributions to simulate data with less obvious spatial structures. For these simulations, all the methods have similar performance as simulation 1 but show different adaptations. For instance, comparing simulation 2 with simulation 1, BASIN achieves 0.7\% higher SSIM while RMSE and JSD are almost unchanged. SPOTlight and CARD gain obvious improvement in all of the three metrics, and SpatialDWLS performs worse on the three metrics. Overall, BASIN still performs better than other methods for all the simulations. On the other hand, Stereoscope and SPOTlight have relatively larger errors in acinar and cancer regions (Supplementary Fig.~\ref{sim_diff}). SpatialDWLS has relatively larger errors in ductal region, and RCTD has relatively larger errors in acinar and ductal regions (Supplementary Fig.~\ref{sim_diff}). 

By comparing the outputs with the ground truth (Supplementary Fig.~\ref{sim_maps}), one can find that SpatialDWLS and RCTD tend to make each spot dominated with single cell type, which is also reflected in the results of the real data later. Although such a property can be helpful to distinguish different regions, it generates unrealistic proportions when there are multiple cell types in one spot. In contrast, Stereoscope and SPOTlight tend to generate more uniformly distributed results, making it difficult to distinguish different regions when the spatial structure is complicated (for example, the results of simulation 4 in Supplementary Fig.~\ref{sim_maps}). CARD and BASIN tend to generate proportions that match the best data, while BASIN has fewer errors overall (Supplementary Fig.~\ref{sim_diff}).

\subsection{Application in human pancreatic ductal adenocarcinoma data A (PDAC-A)}

\begin{figure}[htbp]
\centering
\includegraphics[width=0.9\textwidth]{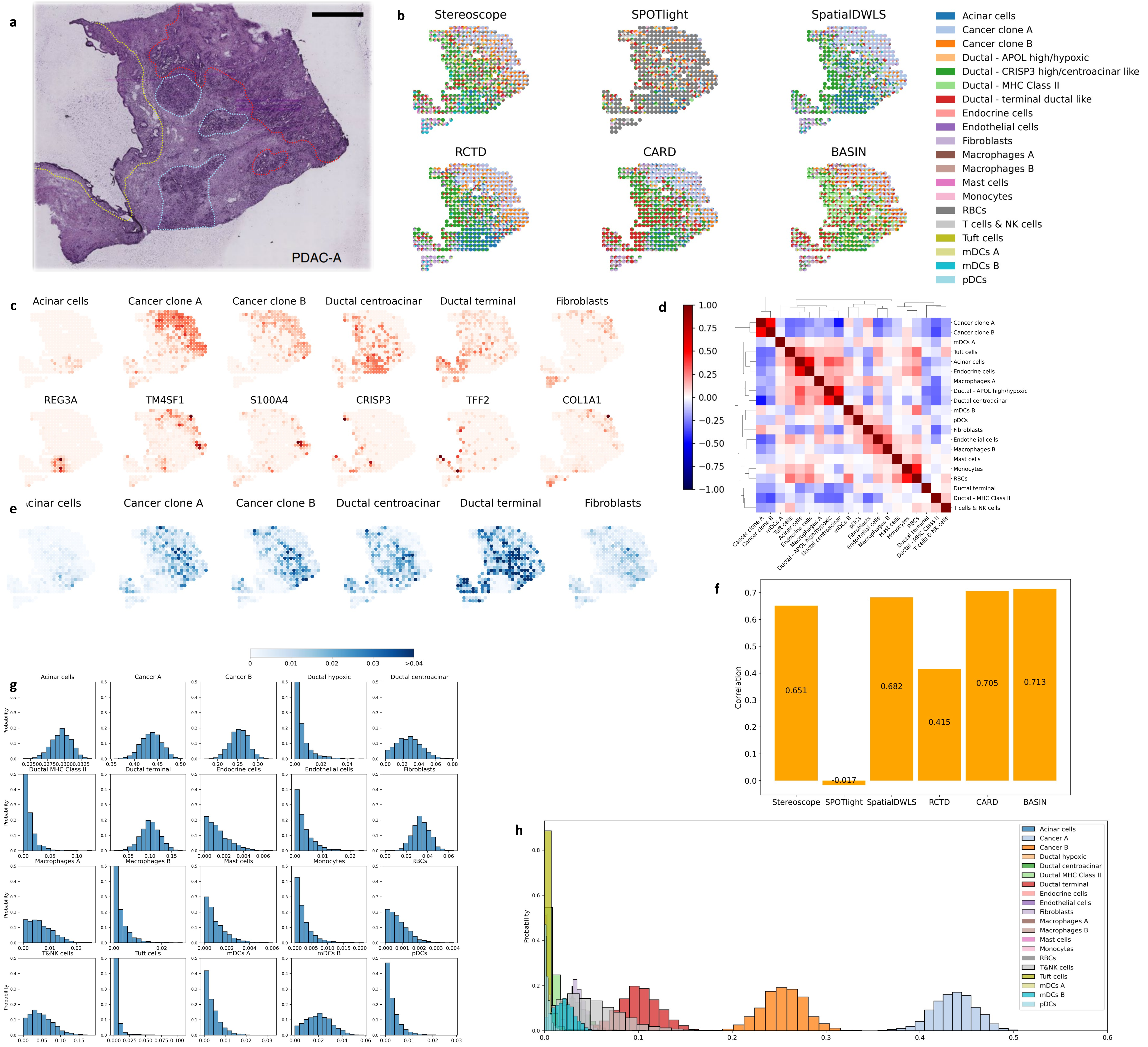}
\caption{Application in PDAC-A data. \textbf{a}, The histological image \cite{moncada2020integrating} with annotations of four regions: cancer (red), pancreatic (blue), ductal (yellow) regions and interstitium (the rest). \textbf{b}, The scatter pie plot of the outputs showing the proportions of the twenty cell types at each spot. We compare BASIN with other five deconvolution methods. \textbf{c}, The first row shows the proportions of six cell types at each spot. The second row shows the amount of the corresponding marker genes at each spot. \textbf{d}, The clustered correlations between each pair of the cell type proportions across all the spots. \textbf{e}, The standard deviations of the proportions of the six cell types at each spot, calculated with 2000 samples. \textbf{f}, The correlations between the predicted cell type proportions and those of the corresponding scRNA-seq data. \textbf{g}, The  histograms of the cell type proportions at one spot selected in the cancer region, calculated with 2000 samples. \textbf{h}, The histograms are plotted together.}\label{pdaca}
\end{figure}

We also evaluated BASIN with real-world spatial transcriptomics and compared with other methods. The accessing and processing of these public world data are described in Method. We first evaluate BASIN with human pancreatic ductal adenocarcinoma (PDAC) data \cite{moncada2020integrating} and utilize scRNA-seq data with twenty cell types generated from the same resource. We tested both tumor cryosections PDAC-A and PDAC-B. The H\&E staining image of PDAC-A in Fig.~\ref{pdaca}a can be roughly divided into four regions: cancer cells (red), duct epithelium (yellow), pancreatic tissue (blue) and stroma (the rest area). There is a clear difference in the proportions of cell types among the cander region, duct epithelium and pancreatic tissue while the stroma is similar to the duct epithelium \cite{moncada2020integrating}.

In Fig.~\ref{pdaca}b we show the cell type proportions of each spot and compare with other five deconvolution methods. Among the six methods, SPOTlight fails to distinguish the cancer regions, while the other five methods clearly show the cancer region dominant with two kinds of cancer cells. According to the scRNA-seq data from the same resource, both the two cancer clusters are abundant in the cancer region, which is reflected in the results of Stereoscope, RCTD and BASIN but not in SpatialDWLS and CARD. Besides, only BASIN shows the existence of ductal terminal cells in the cancer region \cite{moncada2020integrating}. For non-cancer regions, SpatialSWLS, RCTD, CARD and BASIN correctly distinguish the stroma and pancreatic tissue. Only CARD and BASIN reflect that centroacinar cells and terminal duct cells exist at the junction between peripheral acinar cells and the adjacent ductal epithelium \cite{rovira2010isolation}. However, only BASIN predicts a significant amount of antigen-presenting ductal cells expressing  MHC Class II near the duct epithelium, matching the scRNA-seq data. To further confirm the accuracy of BASIN, in Fig.~\ref{pdaca}c we compare the distribution of the six cell types having high enrichment in certain regions with corresponding marker genes \cite{moncada2020integrating}. Although the distribution of cells can be affected by many critical genes rather than a single one, the marker genes can be a reasonable reference to show the accuracy if they match well with the cells. 

In Fig.~\ref{pdaca}d we show the correlations of cell type proportions across the spatial locations as a hierarchically-clustered heatmap. It shows the spatial correlations for some pairs of cell types inferred by BASIN. Specifically, the two cancer clusters are distinguished from other cells spatially. Acinar and endocrine cells have higher spatial correlation because they have high enrichment only in the pancreatic tissue. On the other hand, cancer and centroacinar cells show a relatively low spatial correlation, matching the fact that centroacinar cells only have high enrichment in the stroma and duct epithelium \cite{moncada2020integrating}. In Fig.~\ref{pdaca}e we show the derived standard deviations of the proportions of the six cell types, calculated with 2000 posterior draws. It indicates how confident we can be in the estimates and the range of plausible values. For instance, the standard deviations of the proportions are lower in the cancer region and pancreatic tissue, but higher in the central region where different regions are mixed (considering that the proportions are ranged from 0 to 1). That means the central region is more difficult to distinguish, as shown in the histology image and the results in Fig.~\ref{pdaca}b.

Since the scRNA-seq data are obtained in the same tissue as the spatial transcriptomics data, in Fig.~\ref{pdaca}f we evaluate the Pearson product-moment correlation between the overall mean cell type proportions of the compared methods and that in the scRNA-seq data, showing that stereoscope, SpatialDWLS, CARD and BASIN obtain more accurate overall mean cell type proportions while BASIN performs the best. In Fig.~\ref{pdaca}g and h, we select a spot from the cancer region and plot histograms of the estimated cell type proportions based on 2,000 posterior samples. All twenty cell types follow truncated normal distributions with varying means and standard deviations. As expected, the two cancer clones are predominant at this spot, further confirming the accuracy of our method.

\subsection{Application in human pancreatic ductal adenocarcinoma data B (PDAC-B)}

\begin{figure}[htbp]
\centering
\includegraphics[width=0.9\textwidth]{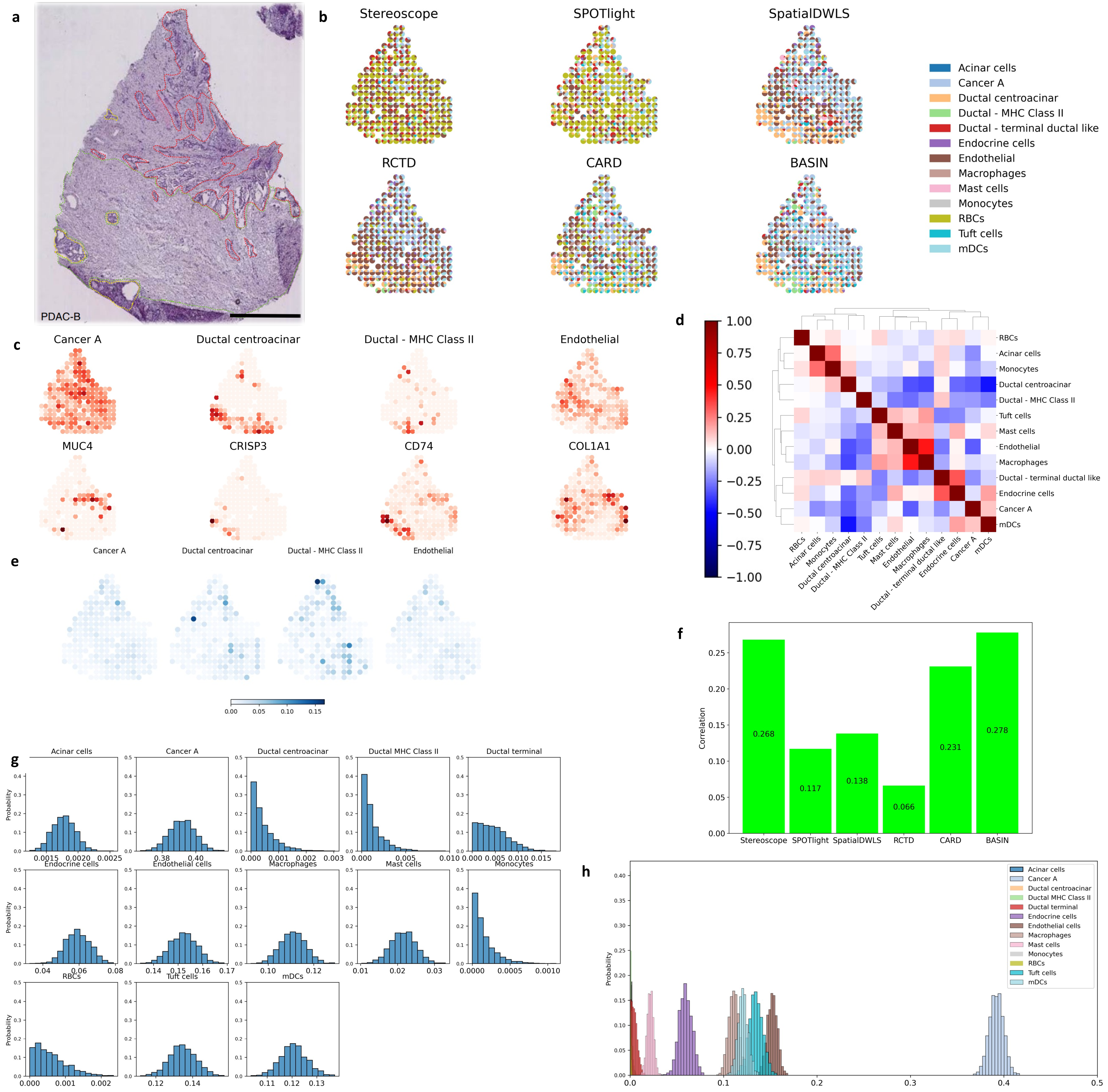}
\caption{Application in PDAC-B data. \textbf{a}, The histological image \cite{moncada2020integrating} with annotations of three regions: cancer (red), ductal (yellow) regions and interstitium (green). \textbf{b}, The scatter pie plot of the outputs showing the proportions of the thirteen cell types at each spot. We compare BASIN with other five deconvolution methods. \textbf{c}, The first row shows the proportions of three cell types at each spot. The second row shows the amount of the corresponding marker genes at each spot. \textbf{d}, The clustered correlations between each pair of the cell type proportions across all the spots. \textbf{e}, The standard deviations of the proportions of the three cell types at each spot, calculated with 2000 samples. \textbf{f}, The correlations between the predicted cell type proportions and those of the corresponding scRNA-seq data. \textbf{g}, The  histograms of the cell type proportions at one spot selected in the cancer region, calculated with 2000 samples. \textbf{h}, The histograms are plotted together.}\label{pdacb}
\end{figure}

In this section we evaluate the performance of our method on PDAC-B that is from a different patient than of PDAC-A data. The H\&E staining image of PDAC-B in Fig.~\ref{pdacb}a shows annotations of three regions: cancer region (red), duct epithelium (yellow) and interstitium (green), corresponding to thirteen cell types in its scRNA-seq data. In this data, each region has one or several dominant cell types \cite{moncada2020integrating}: cancer cluster A (red), ductal cells (yellow) and endothelial cells (green). Because these cell types are also the most abundant ones in the scRNA-seq data from the same resource, we mainly consider them in this section.

We compare the predicted proportions of BASIN with other five deconvolution methods in Fig.~\ref{pdacb}b, showing that our proposed method can best match each region with its correct dominant cell types. Among the six methods, Stereoscope and SPOTlight fail to distinguish the cancer and ductal cells. CARD generates excessively high proportion of RBCs (red blood cells) and fails to reflect the region of duct epithelium clearly. RCTD identifies the cancer region and duct epithelium, but the proportion of endothelial cells in the duct epithelium is too high. SpatialDWLS and our method can distinguish the three regions correctly, while BASIN can also distinguish ductal cells expressing major histocompatibility complex (MHC) class II from ductal centroacinar cells within the duct epithelium region. In Fig.~\ref{pdacb}c we show that the distributions of the four cell types match well with their highly expressed genes as well as the histology image. Specifically, the distribution of ductal cells with MHC class II corresponds to that of \textit{CD74}, a MHC class II gene \cite{moncada2020integrating}. In Fig.~\ref{pdacb}d, we show the correlations of cell type proportions across the spatial locations as a hierarchically-clustered heatmap. The endothelial, ductal centroacinar, cancer cells and ductal MHC class II are divided into different clusters according to the heatmap. In Fig.~\ref{pdacb}e we show the derived standard deviations of the proportions of these four cell types. It shows that the top and lower-right parts of the tissue have relatively higher uncertainty, where different regions are mixed together. In the histology image Fig.~\ref{pdacb}a, it corresponds to the left and bottom boundary of the cancer region. 

Similar to PDAC-A, in Fig.~\ref{pdacb}f, we evaluate the correlation between the overall mean cell type proportions of the compared methods and that in the scRNA-seq data. Since all the compared methods fail to distinguish the terminal ductal cells, the cell type correlations are lower than 0.3 overall. But it still show that Stereoscope and BASIN obtain the most accurate overall mean cell type proportions. In Fig.~\ref{pdacb}g and Fig.~\ref{pdacb}h, we select a spot in the cancer region and plot the histograms of 2000 samples separately and together. All the thirteen cell types follow the truncated normal distributions, and the cancer cell is the most abundant at this spot.

\subsection{Application in mouse olfactory bulb (MOB) data}

\begin{figure}[h]
\centering
\includegraphics[width=0.9\textwidth]{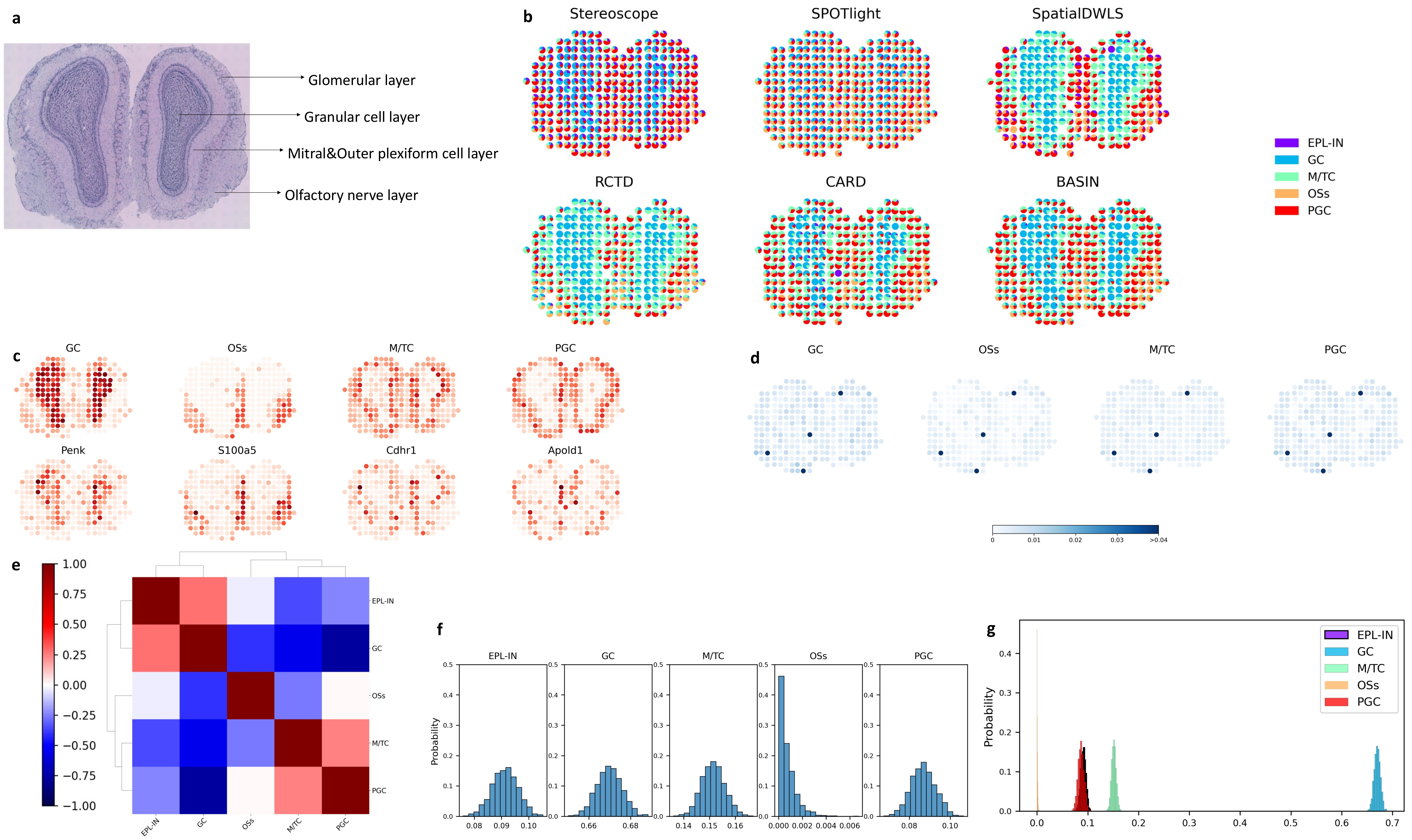}
\caption{Application in MOB data. \textbf{a}, The histological image from \cite{staahl2016visualization} clearly shows a spatial structure with different layers. \textbf{b}, The scatter pie plot of the outputs showing the proportions of the five cell types at each spot. We compare BASIN with other five deconvolution methods. \textbf{c}, The first row shows the proportions of four cell types at each spot. The second row shows the amount of the corresponding marker genes at each spot. \textbf{d}, The standard deviations of the proportions of the five cell types at each spot, calculated with 2000 samples. \textbf{e}, The clustered correlations between each pair of the cell type proportions across all the spots. \textbf{f}, The  histograms of the cell type proportions at one spot selected in the granular cell layer, calculated with 2000 samples. \textbf{g}, The histograms are plotted together.}\label{mob}
\end{figure}

We evaluate the replicate 12 of mouse olfactory bulb (MOB) data from \cite{staahl2016visualization} and utilize the scRNA-seq data of GSE121891 \cite{tepe2018single} as the reference. The H\&E staining image in Fig.~\ref{mob}a shows an obvious layered structure dividing the tissue into four regions: granule cell layer, mitral cell and outer plexiform layer, glomerular layer, and olfactory nerve layer. And there are five groups of cells in the scRNA-seq data: external plexiform layer interneurons (EPL-IN), granule cells (GC), mitral/tufted cells (M/TC), olfactory sensory neurons (OSs) and periglomerular cells (PGC). Except EPL-IN, the other four groups of cells in the scRNA-seq data correspond to the four layers in the ST data and the deconvolution results should reflect such structures too. 

We compare BASIN with other five deconvolution methods in Fig.~\ref{mob}b, which shows that our method predicts the most accurate cell-type proportions. In comparison,  Stereoscope and SPOTlight predict the correct structure and but the effect of spatial correlation is too strong and the cell type proportions are inaccurate. For SpatialDWLS and RCTD also show the layers clearly but fail to reflect the complete PGC layer. Besides, SpatialDWLS shows blurred boundaries of GC, M/TC and OSs, and the overall proportion of MT/C in RCTD is too dominant. The result of CARD and BASIN are overall the best, both of which identify the M/TC and PGC layers accurately, while CARD shows more PGCs in the GC layer. Since EPL-in is trivial in the results, we compare the distribution of the other four cell types with corresponding marker genes in Fig.~\ref{mob}c and show that they match well with each other. Both the cell type proportions and marker genes reflect the correct structures for each layer and show obvious boundaries against each other. 

In Fig.~\ref{mob}d we show the standard deviations of each cell type at each location to reflect the uncertainty of our results. Obviously, there are four spots with much higher standard deviations, indicating that they are difficult to deconvolve. For example, the upper right spot is predicted as 100\% EPL-IN by SpatialDWLS, and the middle one is predicted as 100\% EPL-IN by CARD. Besides, the spot at bottom is removed by SPOTlight, SpatialDWLS, RCTD and CARD because none of the cell types match the gene profile at this spot. Therefore, our method can help find such highly ambiguous spots, which may be due to technical artifacts or complex cellular interactions. Fig.~\ref{mob}e reflects that GC, OSs and M/TC layers are identified explicitly. M/TC and PGC have higher correlation because these two layers are adjacent and harder to distinguish. In Fig.~\ref{mob}f and Fig.~\ref{mob}g, we select a spot in the granule cell layer and plot the histograms of 2000 posterior samples separately and together, showing that the GC (granule cell) is predominant at this spot. The EPL-IN, GC, M/TC and PGC are distributed as normal distributions obviously. The OSs is shown as a truncated normal distribution because this cell type is trivial in this region. These results confirm the accuracy of our method and prove that BASIN can output cell type proportions in terms of truncated normal distributions. Some look normal distributed because the tail probability is very close to zero, therefore we didn't draw any sample from there.

\subsection{Application in seqFISH+ mouse cortex data}

\begin{figure}[h]
\centering
\includegraphics[width=0.9\textwidth]{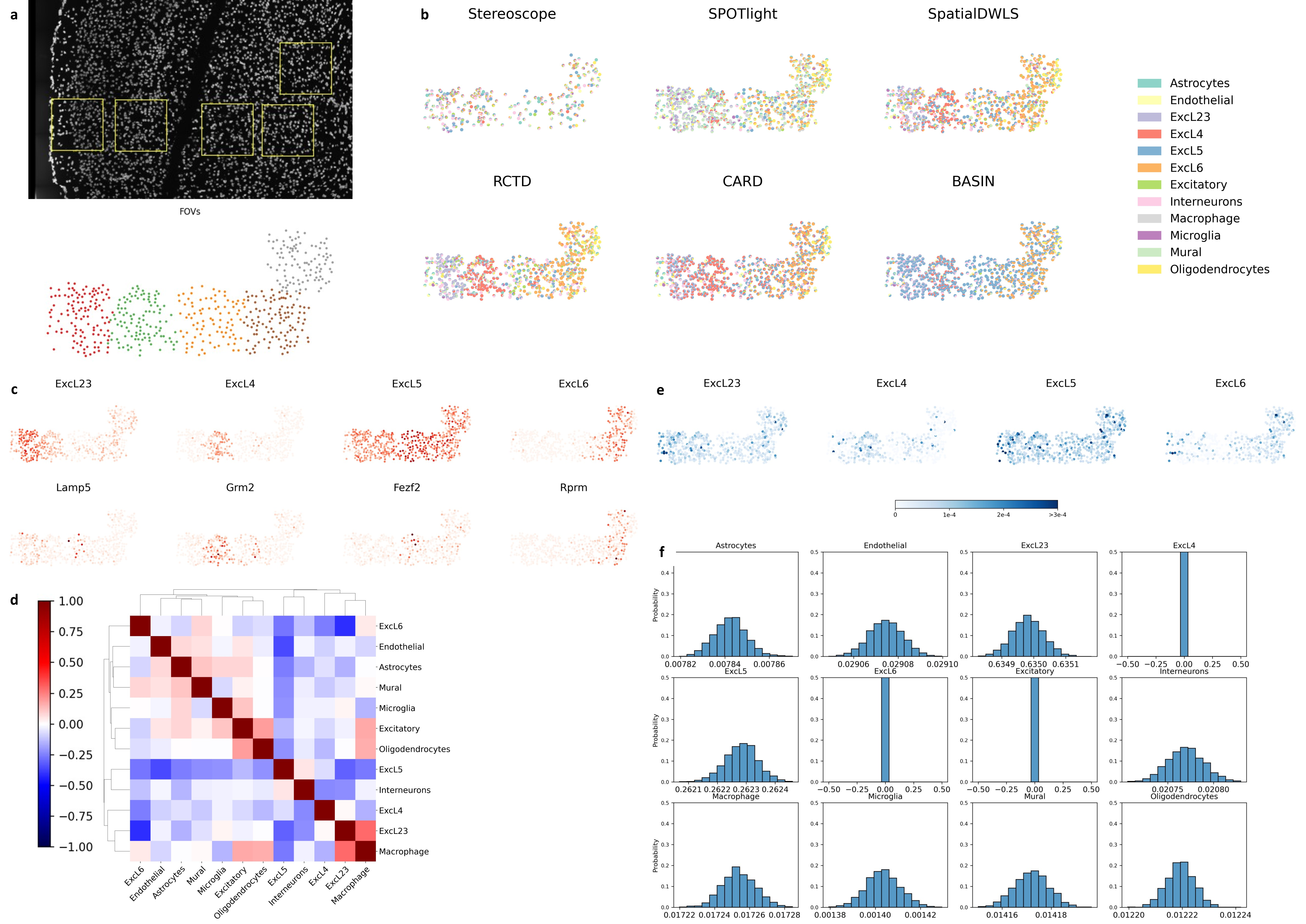}
\caption{Application in seqFISH+ data. \textbf{a}, Upper: The spatial locations of the cells are from five field of views (FOVs) on the mouse cortex \cite{eng2019transcriptome}. Lower: The five fields are stitched together with different x and y-offset values for each FOV. The five FOVs are represented with different colors. \textbf{b}, The scatter plot of the outputs showing the dominant cell type of each spot. \textbf{c}, The first row shows the proportions of four excitatory clusters at each spot. The second row shows the amount of the corresponding marker genes at each spot. \textbf{d}, The clustered correlations between each pair of the cell type proportions across all the spots. \textbf{e}, The standard deviations of the proportions of six cell types at each spot, calculated with 2000 samples. \textbf{f}, The probability histograms of the cell type proportions at one spot selected in the granular cell layer, calculated with 2000 samples. Since the standard deviations are too small, they are not plotted together as other data.}\label{seqfish}
\end{figure}

We evaluated the performance in seqFISH + cortex data \cite{eng2019transcriptome} and used GSE102827 scRNA-seq data \cite{hrvatin2018single} as a reference. The data is generated from five field of views (FOVs) of the cortex, as shown in Fig.~\ref{seqfish}a, and the coordinates of the spots within each FOV are adjusted so that the five FOVs are assembled together. The FOVs correspond to different excitatory neuron layers (ExcLs) with different gene expressions. From the left to right, the first three FOVs correspond to ExcL23, ExcL4 and ExcL5, and the last two FOVs are dominant with ExcL6.

BASIN clearly shows the layered structures of the cortex in Fig.~\ref{seqfish}b and c which match with the layers in Fig.~\ref{seqfish}a. In Fig.~\ref{seqfish}d, one can also observe that the ExcLs predicted by BASIN are located in different clusters with low spatial correlations with each other. By contrast, Sterescope does not show the layers clearly. SPOTlight, SpatialDWLS, RCTD and CARD identify the layers of ExcL23 and ExcL4 well but fail to distinguish the layers of ExcL5 and ExcL6. In our results, the predicted ExcL5 has high proportions in general,l, which is also reflected in Fig.~\ref{seqfish}e where the ExcL5 proportions have higher standard deviations than the others. All these results show that ExcL5 is relatively harder to distinguish. Finally, Fig.~\ref{seqfish}f shows the probability histogram of the proportions of cells of a spot selected in the ExcL23 layer, and ExcL23 cells take 63. 5\% on average for the 2000 posterior samples. The standard deviations are close to zero, indicating low gene similarities among cell types. One should note that the proportions of ExcL4, ExcL6, and other excitatory cells appear as small non-zero values due to the smoothing effect of the histogram's bandwidth setting.

\section{Discussion}\label{discussion}
In this paper, we have presented BASIN, a cellular deconvolution method for spatial transcriptomics that integrates gene expression information from scRNA-seq data, spatial locations, and histology images to estimate cell-type proportions at each location. Compared with existing approaches, BASIN demonstrates superior performance in multiple simulation settings and real spatial transcriptomics datasets generated from diverse tissues and technologies, highlighting its robustness and ability to preserve spatial structure. In the PDAC data, BASIN shows clear boundaries of subregions on the tissue and accurately distinguishes different cell types within each region. In the MOB data, it clearly shows the layer structures. In the seqFISH+ mouse cortex data, it also predicts different excitatory layers correctly. Despite involving sampling from high-dimensional probability distributions, BASIN is highly efficient (Supplementary Fig.~\ref{time}) and outperforms most existing methods in computation time.

What makes BASIN unique is that it generates results from a probability distribution, which enables easy parameter tuning and uncertainty quantification. We propose a matrix variate Bayesian method for the NMF problem and combine with graph Laplacian and sparsity priors. Results in different data show that the proportions at each location follow truncated normal distribution as we assume with different uncertainties. We report the standard deviations of the estimated cell type proportions at each spot to quantify the uncertainty arising from gene expression similarity among cell types in the scRNA-seq data, the spatial structure in the transcriptomics data, and the selection of hyperparameters.

Cellular deconvolution methods like BASIN still have room to be improved. First, the histology images that contain spatial information could be better utilized.  The histology images are typically not aligned with the ST and have much higher spatial resolutions than ST. In our method, we align the spot coordinates in ST data with the histology image manually and resize the image to a smaller size like $1024\times 1024$, and we compute the average intensity of a small square area ($5\times 5$) around each spot location in the ST data. However, better image alignment techniques could be studied to align the two sets of data accurately and further increase the accuracy. Second, sampling from truncated multivariate normal distributions in our method may still fail when the sampling dimension is too high. The issues arise either from the sampling algorithm itself which can be addressed through resampling or from inappropriate hyperparameter choices that render the covariance matrices to be non-invertible. Although the current sampling method \cite{li2015efficient} is the best for BASIN considering both efficiency and accuracy, better substitute can be used in the future. 

BASIN can be extended in several ways in future work. First, while BASIN incorporates gene similarity from both ST and scRNA-seq data, it currently treats all genes equally. However, gene importance varies: some genes are uniformly expressed across the tissue, while others may be uninformative or even misleading for distinguishing cell types. The gene expressions can be weighted by their contribution of differentiating cell types and a weight matrix can be added to the NMF-based models. Second, BASIN is a reference-based method using scRNA-seq data and can also be extended to a reference-free method. We provide a matrix variate method for Bayesian NMF but one factorized matrix is known from scRNA-seq data. If scRNA-seq reference is not available, one needs to derive the posteriors for both of the factorized matrices. In the future, we should think of how to use ST data to help cluster the cell-types across spatical locations \cite{zhao2021spatial, lopez2022destvi}. Third, BASIN can be generalized to tensor methods by involving tensor normal distributions \cite{manceur2013maximum}, enabling us to deal with spatial dimensions separately and extract meaningful latent structures from high dimensional ST data \cite{burch2025towards}. Combination with tensor decomposition techniques may be needed to reduce computational consumption \cite{song2023gntd, li2021imputation, broadbent2024deciphering}.

\section{Methods}
\subsection{The BASIN method}
The goal of cell type deconvolution is to estimate the proportions of different cell types at each spot of the ST data. Suppose the ST data comprises expressions of $n$ genes across $p$ locations, and the reference scRNA-seq data contains expressions of the same $n$ genes across cells annotated into $c$ distinct cell types. Our objective is to estimate the proportions of these $c$ cell types at each of the $p$ spatial locations. In ST data, we denote $\mathbf{X}$ as the $n\times p$ gene expression matrix, where $n$ informative genes are measured across $p$ spatial locations. From the scRNA-seq data, we derive a matrix $\mathbf{B}$ representing the $n\times c$ cell-type specific expression matrix for the same set of $n$ informative genes. Each element denotes the mean expression level of an informative gene in a specific cell type. 

We introduce $\mathbf{V}$ as the $c\times p$ cell type composition matrix, where each column of $\mathbf{V}$ indicates the proportions of the $c$ cell types at each spatial location. Our objective is to estimate $\mathbf{V}$ using both $\mathbf{X}$ (the ST data) and $\mathbf{B}$ (derived from the scRNA-seq data). Naturally, the matrix $\mathbf{V}$ is constrained to be non-negative. Besides, since most cell types at each spot take little proportions, to increase the difference among cell type proportions within one spot to better distinguish different regions, we add a sparsity prior to $\mathbf{V}$ to keep $\mathbf{V}$ getting too uniformly distributed. Laplacian distribution (also called double exponential distribution) is a Bayesian equivalent of $l_1$ norm regularization \cite{garrigues2010group} and is applied to induce sparsity in Bayesian models \cite{costa2015sparse}. To induce non-negativity, we consider the positive part of Laplacian distribution and therefore the prior becomes an exponential distribution:
\begin{align}
& \mathbf{V}_{ij} | \eta \sim \text{Exp}(\eta)
\end{align}
where $\mathbf{V}_{ij}\ge 0$ for $1\le i\le c$ and $1\le j\le p$.
We represent it in a matrix form to combine with the Equation~(\ref{likelihood}) and make an ablation study to show the effect of the exponential distribution in Supplementary Fig.\ref{sparse}. We utilize a non-negative matrix factorization (NMF) model to link these matrices:
\begin{align}
\mathbf{X} = \mathbf{BV} + \mathbf{E}
\end{align}
Since $\mathbf{X}$ is integer data, we transform it to continuous-valued data with 'lop1p' and assume the error matrix $\mathbf{E}$ as a Laplacian-structured Gaussian Markov Random Fields \cite{ying2021minimax} whose precision matrix (inverse of covariance matrix) is the graph Laplacian:
\begin{align}
\mathbf{E} \sim \mathcal{MN}(0, \sigma^2 \mathbf{I_n}, \mathbf{L}^{-1})
\end{align}
where $\sigma^2$ is a parameter controlling the scale of the error, $\mathbf{I_n}$ is an identity matrix of size $n\times n$ and $\mathbf{L}$ is a $p\times p$ matrix representing the graph Laplacian of the spatial domain. $\mathbf{L}$ is defined as $\mathbf{L}=\mathbf{D}-\mathbf{A}$ where $\mathbf{A}$ is the weight (or adjacent) matrix and $\mathbf{D}$ is a diagonal matrix with  $\mathbf{D}_{ii} = \sum_j \mathbf{A}_{ij}$. There are different choices for the definition of weight matrix $\mathbf{A}$, e.g., 0-1 weighting, Gaussian kernel weighting and dot-product weighting \cite{cai2010graph}. We use Gaussian kernel weighting because it can measure distance between two nodes and is naturally suitable for a 2D spatial domain. We also consider the intensity difference of the aligned coordinates on the histology image (if available). Therefore, the weight matrix is defined as:
\begin{align}
	\mathbf{A}_{ij} =
	\begin{cases}
		\exp(-\frac{||s_i - s_j||^2 + \mu ||I_i-I_j||^2}{\sigma_A^2}) & \text{if } i \neq j, \\
		0 & \text{if } i = j,
	\end{cases}
\end{align}
where $||s_i - s_j||^2$ is the Euclidean distance between spots $i$ and $j$, and $||I_i-I_j||^2$ is the intensity difference obtained from the histology image. $\mu$ and $\sigma^2_A$ are constants controlling the weight of intensity difference and adapting spatial coordinates with different magnitudes. By setting $\mathbf{L}^{-1}$ as the column covariance matrix, we assume that the covariance of two spatially closer locations is larger, based on the fact that neighboring spatial locations have more similar gene expression \cite{ma2022spatially}. Give above, $\mathbf{X}$ is also following matrix normal distribution:
\begin{align}
\mathbf{X} | \mathbf{V}, \sigma^2 \sim \mathcal{MN}(\mathbf{BV}, \sigma^2 \mathbf{I_n}, \mathbf{L}^{-1})
\label{likelihood}
\end{align}
The priors of the hyperparameters $\sigma^2$ and $\eta$ also need to be defined. The choice of prior depends on the conjugacy between the prior and the likelihood of the parameter. Therefore we set their priors as follows:
\begin{align}
& \sigma^2 \sim \text{Inverse-Gamma}(a_1, b_1)  \\
& \eta \sim \text{Gamma}(a_2, b_2) 
\end{align}
where $a_1$, $a_2$, $b_1$ and $b_2$ are hyperparameters.

\subsubsection{Posteriors}
By applying Bayes' theorem, the conditional posterior distribution of $\mathbf{V}$ can be calculated by the formula:
\begin{align}
P(\mathbf{V}|\mathbf{X}) & \propto P(\mathbf{X}|\mathbf{V})P(\mathbf{V}) 
\end{align}
Combining the non-negativity of $\mathbf{V}$ brought by the exponential distributions in the prior, We show that the posterior of $\mathbf{V}$ is a truncated matrix normal (TMN) distribution:
\begin{align}
\mathbf{V} | \mathbf{X} \sim \mathcal{TMN}
((\mathbf{B}^T\mathbf{B})^{-1}(\mathbf{B}^T\mathbf{X} - \sigma^2 \eta \mathbf{J_V} \mathbf{L}^{-1}), 
\; (\mathbf{B}^T\mathbf{B})^{-1} \sigma^2, 
\;  \mathbf{L}^{-1}) \quad (\mathbf{V}\ge 0)
\label{posterior_v}
\end{align}
where $\mathbf{J_V}$ is an ``all ones matrix" of the same dimension as $\mathbf{V}$. Specifically, the mean matrix $(\mathbf{B}^T\mathbf{B})^{-1}(\mathbf{B}^T\mathbf{X} - \sigma^2 \eta \mathbf{J_V} \mathbf{L}^{-1})$ captures the cell type proportions at each location. The column covariance $\mathbf{L}^{-1}$ (spatial dimension) still involves the spatial structure defined by the graph Laplacian, while the row covariance $(\mathbf{B}^T\mathbf{B})^{-1} \sigma^2$ (cell type dimension) is determined by the cosine similarity of the cell types measured by $\mathbf{B}^T\mathbf{B}$. 

Likewise, the conditional posterior distributions of $\sigma^2$ and $\eta$ are calculated as:
\begin{align}
& \sigma^2 | \mathbf{X},\mathbf{V} \sim \text{Inverse-Gamma}(a_1 + np/2 + cp/2, \quad b_1 + \frac{\text{tr}(\mathbf{LE}^T \mathbf{E})}{2} )) 
\label{posterior_sigma}
\end{align}
\begin{align}
& \eta | \mathbf{V} \sim \text{Gamma}(a_2 + cp, \quad 1/(\text{tr}(\mathbf{J_V}^T \mathbf{V})) + \frac{1}{b_2})))
\label{posterior_eta}
\end{align}
When choosing the hyperparameters, we want to reduce the influence of them. We set $a_{1,2}$, $b_1$ and $\frac{1}{b_2}$ to be really small, e.g.,$10^{-4}$ to exploit more information from data.

\subsubsection{Overview of the algorithm}
Starting with an arbitrary initial value of $\mathbf{V}$, we iteratively sample the parameters from their conditional posterior distributions using Gibbs sampling. The proposed algorithm is summarized in Algorithm~\ref{algorithm1}.
\begin{algorithm}
\caption{BASIN}\label{algorithm1}
\begin{algorithmic}
\State \textbf{Require:} $\mathbf{X}$, $\mathbf{B}$, $\sigma^2_A$, $\mu$, $\mathbf{L}$, sample size $T$, burn-in period $T_b$ 
\State \textbf{Initialize:} choose $\mathbf{V}^0$
\For{$t$ = 1 to $T$}
    \State Sample $(\sigma^2)^t$ given $\mathbf{V}^{t-1}$
    with Equation~(\ref{posterior_sigma})
    \State Sample $\eta^t$ given $\mathbf{V}^{t-1}$
    with Equation~(\ref{posterior_eta})
    \State Sample $\mathbf{V}^t$ given $(\sigma^2)^t$ and $\eta^t$
    with Equation~(\ref{posterior_v})
    \State Normalize $\mathbf{V}$ so that the column sum is one
\EndFor
\State Discard $\mathbf{V}^1 \cdots \mathbf{V}^{T_b}$
\end{algorithmic}
\end{algorithm}
Unlike general Bayesian NMF models with two unknown matrices, our deconvolution model is reference-based with one unknown matrix. Therefore, in practice, the Gibbs sampler can converge within several steps and the burn-in period can be set within five steps. The value of $\sigma^2_A$ depends on the magnitude of coordinate. For most ST data like MOB and PDAC where the 2D coordinates lie in between $10^1$ and $10^2$, we set $\sigma^2_A=0.1$. For the seqFISH+ data where the coordinates have magnitude between $10^3$ and $10^4$, we set $\sigma^2_A=10^4$. $\mu$ is a hyperparameter controlling the weight of intensity information in the H\&E image. It is set to be one after normalizing image and can be ignored when image is not available.

\subsection{Sampling from truncated matrix normal distribution}
Sampling from a high dimensional truncated matrix normal distribution can be challenging and time consuming. Sampling from matrix normal distribution typically requires first transforming it to a multivariate normal distribution (MVN):
\begin{align}
\mathbf{X} \sim \mathcal{MN}(\mathbf{M}, \mathbf{A}, \mathbf{B}) \quad\to\quad
\text{vec}(\mathbf{X}) \sim \mathcal{MVN}(\text{vec}(\mathbf{M}), \mathbf{B}\otimes \mathbf{A})
\end{align}
and then sample from the multivariate normal distribution. Truncated matrix normal distribution can be sampled in the same way. However, when the dimension gets high, sampling from truncated multivariate normal distribution becomes increasingly difficult because the probabilities of each variate are very small and the acceptance rates are almost zero when using Markov chain Monte Carlo methods. In our case, the sampling dimension is the size of $\mathbf{V}$ ($c\times p$), which is usually in the magnitude of $10^3$ or $10^4$. Therefore, sampling from the posterior distribution of $\mathbf{V}$ consumes more than 90\% time in our algorithm.

Although there are many implementation schemes proposed to sample from truncated MVN \cite{wilhelm2010tmvtnorm, botev2017normal, goodman2010ensemble}, we utilize the sampling method proposed in \cite{li2015efficient} because it's the only one that can deal with high dimension variate according to our test. In short words, it decomposes the truncated multivariate normal distribution into a set of truncated univariate distributions and samples with Gibbs sampling. Although it's efficient, it's still computationally expensive when the parameter dimension is too high (Supplementary Fig.~\ref{time}b). To further improve the efficiency, we simplify the sampling method \cite{li2015efficient} by only considering the truncated range from zero to infinity, while the method originally considers general ranges. Besides, we accelerate the method by removing the unnecessary computations which takes the most time in sampling (details are discussed in Supplementary Information). By simplification, we make the sampling algorithm about 10-1000 times faster than before in our method (Supplementary Fig.~\ref{time}b). We compare the computation times spent on the PDAC-A data in Supplementary Fig.\ref{time}a, showing that BASIN is still efficient when sampling from more than 8000 dimensions.

However, sampling from a a high-dimensional truncated multivariate normal distribution is essentially challenging due to the fact that the probability of each variate is extremely small as the dimension increases. Therefore, in our algorithm, when sampling dimension (number of locations times number of cell types) is larger than $10^4$, we apply rectified matrix normal distribution instead, which is as fast as sampling from regular matrix normal distributions. Specifically, we sample from regular matrix normal distribution first, and then set the negative values to be zeros. We have observed that rectified matrix normal distribution is a good approximation of truncated matrix normal distribution (Supplementary Fig.~\ref{rmn}).

\subsection{Data preprocessing and acquisition}
All the ST data and scRNA-seq data we used are publicly available. Each ST dataset includes two data tables: one containing the gene expression abundance for each spot, and another containing the spatial coordinates (X and Y) of the spots. The scRNA-seq data includes the abundance of each gene for each single cell and a meta data revealing the type of cells. To make sure the two kind of data corresponding well with each other, we preprocess the data using the similar methods as in \cite{Wang2019, ma2022spatially}. First, we filter out the genes not showing in any spots or cells for the two data, and the genes that just show in one of the STdata and scRNA-seq data. Second, the ST and scRNA-seq data are re-ordered by genes, spots, and cells to align perfectly with their corresponding locations and metadata. Third, we construct the cell-type expression matrix using the scRNA-seq data and its meta data, by computing the average gene expression levels for each cell type. Fourth, we calculate the VMR (variance-to-mean ratio) for each gene in each cell type and mean VMR across the cell types. We remove the genes with top 10\% largest mean VMR. Finally, to filter out the genes with low expression heterogeneity and reduce the data size without losing performance, we only keep the genes whose highest mean expression level in one cell type is at least 1.25-log-fold higher than the mean expression level across all remaining cell types.

The mouse olfactory bulb (MOB) data \cite{staahl2016visualization} is downloaded from the website of Spatial Research Lab \\(https://www.spatialresearch.org/). We choose the replicate 12, containing 16034 genes and 282 spots. The corresponding scRNA-seq data is obtained from the website of GSE121891 \cite{tepe2018single}, containing 18560 genes and 12801 cells including 5 cell types. After preprocessing, there are 4284 genes retained. The human pancreatic ductal adenocarcinoma (PDAC) data is downloaded from the website of GSE111672 \cite{moncada2020integrating}, including both ST data and scRNA-seq data generated from the same samples. In the paper we choose PDAC-A ST1 from patient ID GSM3036911, and PDAC-B ST1 from patient ID GSM3405534 as the ST data, and PDAC-A-indrop and PDAC-B-indrop as the scRNA-seq data. The original PDAC-A ST1 data has 19738 genes and 428 spots, and PDAC-A-indrop containing 19738 genes and 1926 cells from 20 cell types. After preprocessing, there are 10431 genes retained. The original PDAC-B ST1 data 19738 genes and 224 spots, and PDAC-B-indrop contains 19738 genes and 1733 cells from 13 cell types. After preprocessing, there are 9401 genes retained. The seqFISH+ mouse cortex data is downloaded from Cai Lab's Github \cite{eng2019transcriptome} (https://github.com/CaiGroup/seqFISH-PLUS), which contains 10000 genes and 524 spots. The corresponding scRNA-seq data is obtained from GSE102827 \cite{hrvatin2018single}, containing 25187 genes and 48266 cells from 8 main types and 33 subtypes. We divide them into 12 types as done in CARD to distinguish different excitatory layers. After preprocessing, there are 9498 genes retained.

\subsection{Simulation strategy}
We use simulated data to evaluate the performance of BASIN and compare with other deconvolution methods. The simulation is based on the real PDAC-A data \cite{moncada2020integrating}. We first divide the spatial domain into three regions according to the histological annotation shown in Supplementary Information. We assume there are three cell types (acinar, cancer cluster 1 and terminal ductal cells) whose proportions follow different probability distributions and are dominant respectively in the three regions. Four simulation schemes are designed as below: 

In simulation 1, each region has a piecewise constant cell type composition in which the dominant cell type has a proportion of 0.7 and the other two types take 0.15. Simulation 2 is based on simulation 1, in which the simulated cell type proportion data is smoothed with a large Gaussian kernel of size $40\times 40$ with $\sigma=0.5$. In simulation 3, the cell type proportions follow Gaussian distributions whose mean values are equal to those of simulation 1 and the standard deviation is 0.05. Then the data is smoothed with the same Gaussian kernel used in simulation 2. In simulation 4, the three cell types follow different Dirichlet distributions in the three regions. For each region. the concentration parameter is equal to 3 for the main type and equal to 1 for the other two types.

For each of the simulations, the simulated cell type proportion matrix is limited to be nonnegative, with each column normalized to sum to one. These simulated $V$'s are treated as ground truth for each of the four simulations and compared  with results of various methods. We extract the gene profiles of these three cell types from the real scRNA-seq data of PDAC-A. Then we obtain the simulated transcriptomic data based on the NMF model $\mathbf{X} = \mathbf{B}\mathbf{V}$ plus a zero mean Gaussian noise with $\sigma=0.5$ and limited to be nonnegative again.

\subsection{Evaluation metrics}
Three metrics are used to evaluate the simulations. Root mean squared error (RMSE) estimates the average errors between the values and their ground truth, defined as 
\begin{align}
\text{RMSE}(\mathbf{a}, \mathbf{b}) = \frac{1}{\sqrt{p}} ||\mathbf{a} - \mathbf{b}||_F
\end{align}
where $p$ is the number of entries of $\mathbf{a}$ and $\mathbf{b}$. Structural similarity index measure (SSIM) is used to measure the similarity between two images. Compared with RMSE , SSIM considers not only absolute errors, but also difference of structural information, defined as
\begin{align}
\text{SSIM}(\mathbf{a},\mathbf{b}) = \frac{(2\mu_a \mu_b + c_1)(2\sigma_{ab} + c_2)}
{(\mu_a^2 + \mu_b^2 + c_1)(\sigma_a^2 + \sigma_b^2 + c_2)}
\end{align}
where $\mu_a$, $\sigma_a^2$ are the mean and variance of $\mathbf{a}$, and $\sigma_{ab}$ is the covariance measuring the structure similarity. $c_1$ and $c_2$ are parameters to stabilize the division with weak denominator. In our case, we compute the average SSIM of each cell type. Jensen-Shannon distance (JSD) measures the difference between two probability distributions and is defined as the square root of Jensen-Shannon divergence. The Jensen–Shannon divergence is defined as:
\begin{align}
\text{JSD}(\mathbf{a},\mathbf{b}) = \frac{1}{2}\text{KL}(\mathbf{a}||\frac{1}{2}(\mathbf{a}+\mathbf{b})) + \frac{1}{2}\text{KL}(\mathbf{b}||\frac{1}{2}(\mathbf{a}+\mathbf{b})) 
\end{align}
where KL means the Kullback–Leibler divergence. For each location, the cell type can be regarded as following a certain distribution. In this case we compute the average JSD of each location.

\subsection{Benchmark methods}
To evaluate the performance of our proposed method, we compared it with several bulk tissue cell-type deconvolution methods using reference scRNA-seq data: Stereoscope, SPOTlight, SpatialDWLS, RCTD and CARD. Stereoscope \cite{andersson2020single} models both the scRNAseq and spatial data using negative binomial distribution. The parameters are estimated by MLE using gradient descent for some steps and then the cell type proportions are obtained with MAP estimate in a similar way. In our analysis, we set the learning rate as 0.01, batch size as 100 and number of steps as 5000 for each data. SPOTlight \cite{elosua2021spotlight} employs a seeded NMF approach to deconvolute ST spots. The process starts by initializing cell-type marker genes and then applies non-negative least squares (NNLS) to further deconvolute the ST capture locations (spots). In both simulation and real data analysis, all genes are used, with a maximum of 75 cells selected per cell type. All other parameters were set to their default values. SpatialDWLS \cite{dong2021spatialdwls} estimates cell proportions by first identifying the cell types likely present at each location. This is accomplished using marker genes through differential expression analysis with Giotto \cite{Chen2023}. It then applies dampened weighted least squares (DWLS) \cite{Tsoucas2019} to infer the fraction of each selected cell type. An enrichment score cutoff of 2 is used to select the cell types, and all other parameters are set to their default values. RCTD \cite{cable2022robust} models the observed gene counts in each pixel of ST data using a Poisson distribution. It links individual cell types with a generalized hierarchical linear regression model, and the cell type proportions are inferred using the maximum-likelihood estimation (MLE) method. For both simulation and real data analysis, we used the ``doublet'' model, which constrains the number of cell types to two per pixel, as recommended in \cite{cable2022robust} to reduce overfitting and increase power. All other parameters were set to their default values. CARD \cite{ma2022spatially} models the the problem as NMF and assumes the cell type composition as a conditional autoregressive (CAR) model. The parameters are estimated with MAP estimation using a NMF framework. In our analysis all hyperparameters were set to their default values.


\section*{Declarations}
\begin{itemize}
\item Funding: Dr. Liangliang Zhang is supported by his junior faculty start-up grant BGT630267, funded by School of Medicine at Case Western Reserve University. Additional support is provided by the Data Management and Statistics Core under grant number 5P30AG072959, led by Dr. Jonathan L. Haines at Cleveland Alzheimer's Disease Research Center.
\item Competing interests: The authors declare no competing interests.
\item Ethics approval and consent to participate: Not applicable.
\item Consent for publication: All authors agree
\item Data availability: All datasets are publicly available.
\item Materials availability: Not applicable.
\item Code availability: https://github.com/Jiasen-Zhang/BASIN-Deconvolution
\item Author contribution: Guo and Zhang provide the main ideas and guide Zhang who implements the ideas numerically and prepares the draft of the manuscript. Qiao helps with comparison with some of the methods. All coauthors contribute to polish the manuscript. 
\end{itemize}
\noindent

\bibliographystyle{unsrt}
\bibliography{ref}

\pagebreak
\begin{center}
\section*{Supplementary Information}
\end{center}

\setcounter{section}{0}
\setcounter{figure}{0}
\setcounter{equation}{0}
\setcounter{algorithm}{0}

\section{Supplementary figures}

\begin{figure}[h]
\centering
\includegraphics[width=0.6\textwidth]{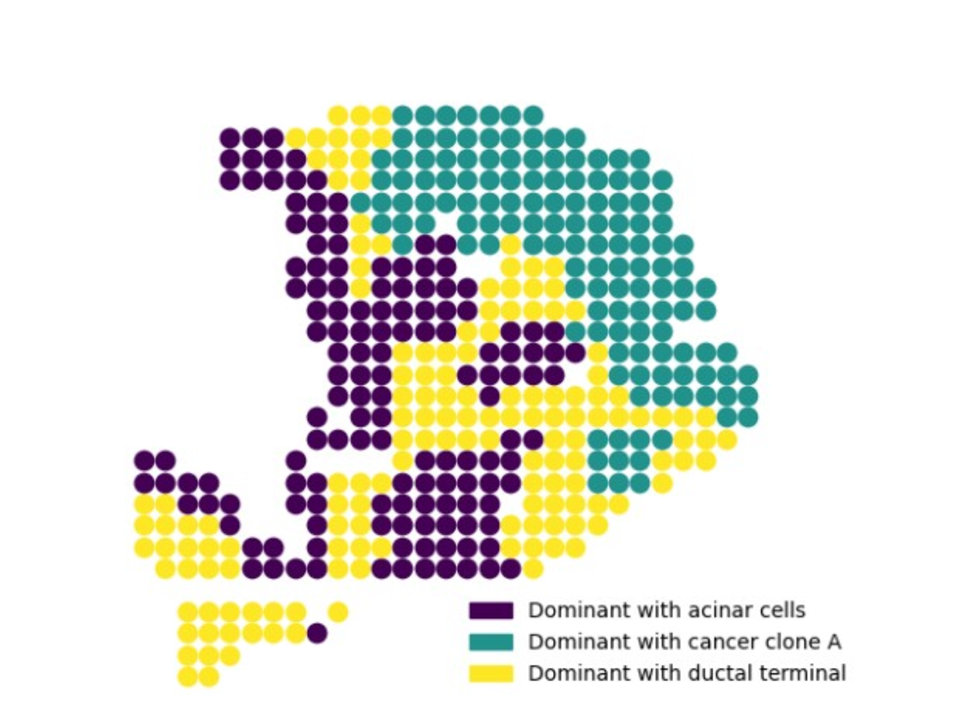}
\caption{Spatial division of our simulated data. The simulation is based on the real PDAC-A data \cite{moncada2020integrating}. We divide the spatial domain into three regions according to the histological annotation, and assume there are three cell types (acinar, cancer cluster A and terminal ductal cells) whose proportions follow different probability distributions and are dominant respectively in the three regions. }\label{sim_regions}
\end{figure}

\begin{figure}[h]
\centering
\includegraphics[width=1.0\textwidth]{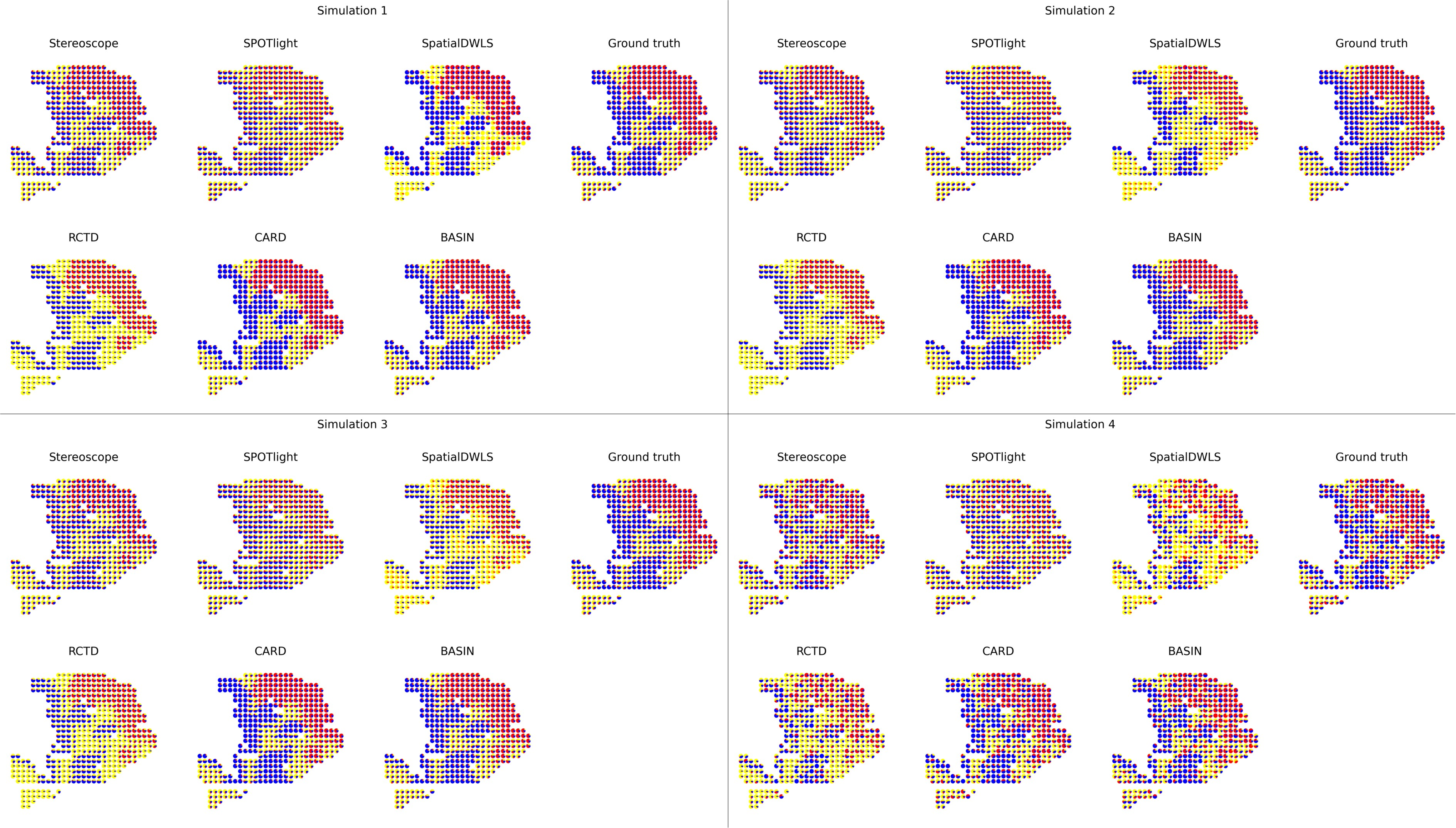}
\caption{The results of the four simulation studies compared with the ground truth. Blue: acinar cells. Yellow: terminal ductal cells. Red: cancer cells.}\label{sim_maps}
\end{figure}

\begin{figure}[h]
\centering
\includegraphics[width=1.0\textwidth]{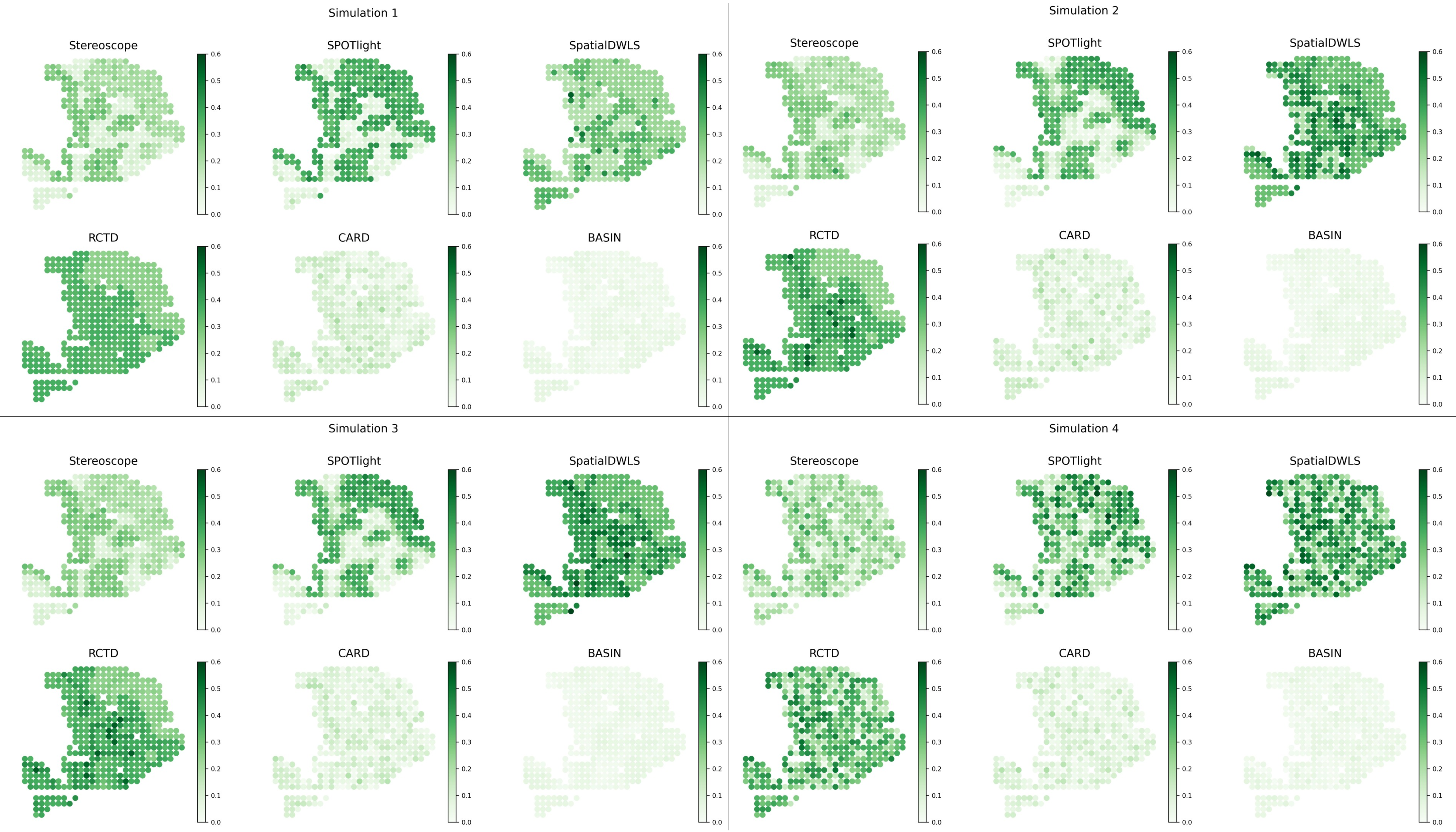}
\caption{The spot-wise RMSE of the four simulations between the six compared methods and the ground truth.}\label{sim_diff}
\end{figure}

\begin{figure}[h]
\centering
\includegraphics[width=1.0\textwidth]{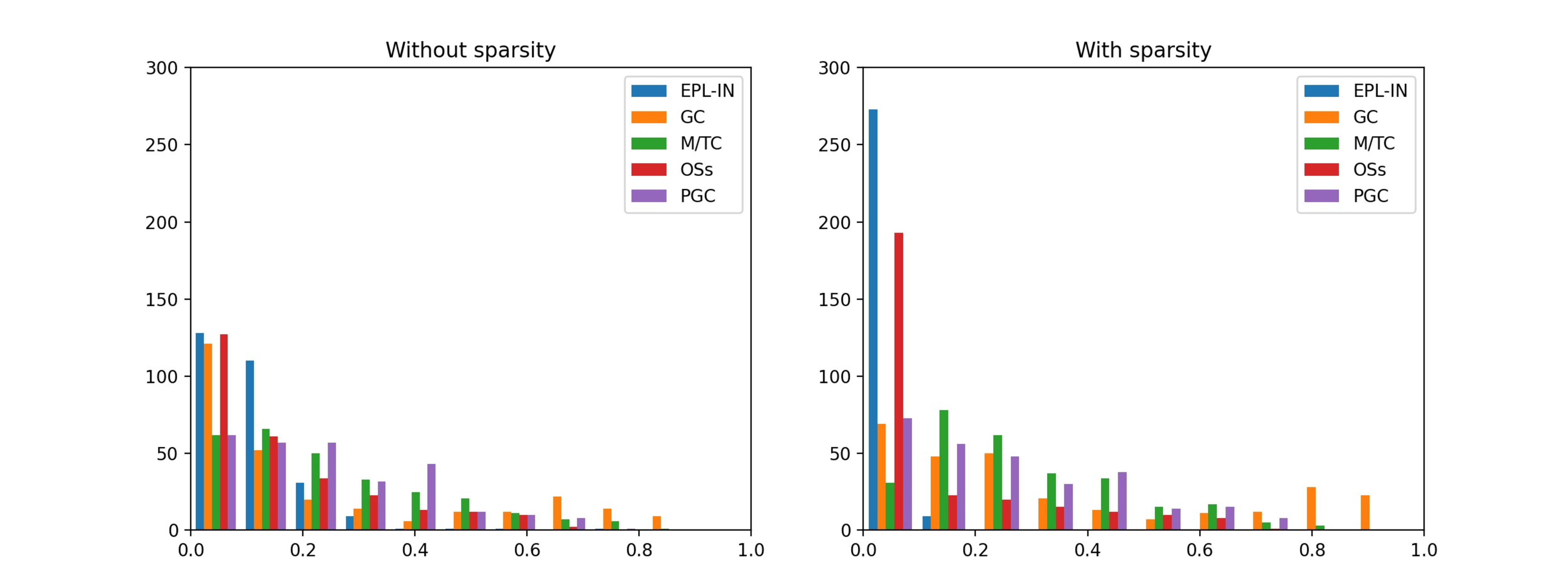}
\caption{We make an experiment to show the effect of the exponential distribution in the prior of $V$. It not only constrains the values of $V$ to be nonnegative, but also introduce sparsity to $V$ to reduce noise and better distinguish the cells. We use the MOB data as an example and plot the histograms of the five cell types without (left) and with (right) the exponential distribution in the prior. We can see that by introducing sparsity, there are more spots with trivial proportion of EPL-IN (blue) and OSs (red) on the right. It matches the fact that EPL-IN is trivial in the tissue and OSs only exist in the a small area. On the other hand, there are more spots with high proportions of GC cells (orange) on the right, which is also reasonable since GC cells mainly distributed in the central dominant area. Therefore, the exponential distribution in the prior help us better identify the dominant cell type of each spot, and avoid getting too uniform distributed results.}\label{sparse}
\end{figure}

\begin{figure}[h]
\centering
\includegraphics[width=1\textwidth]{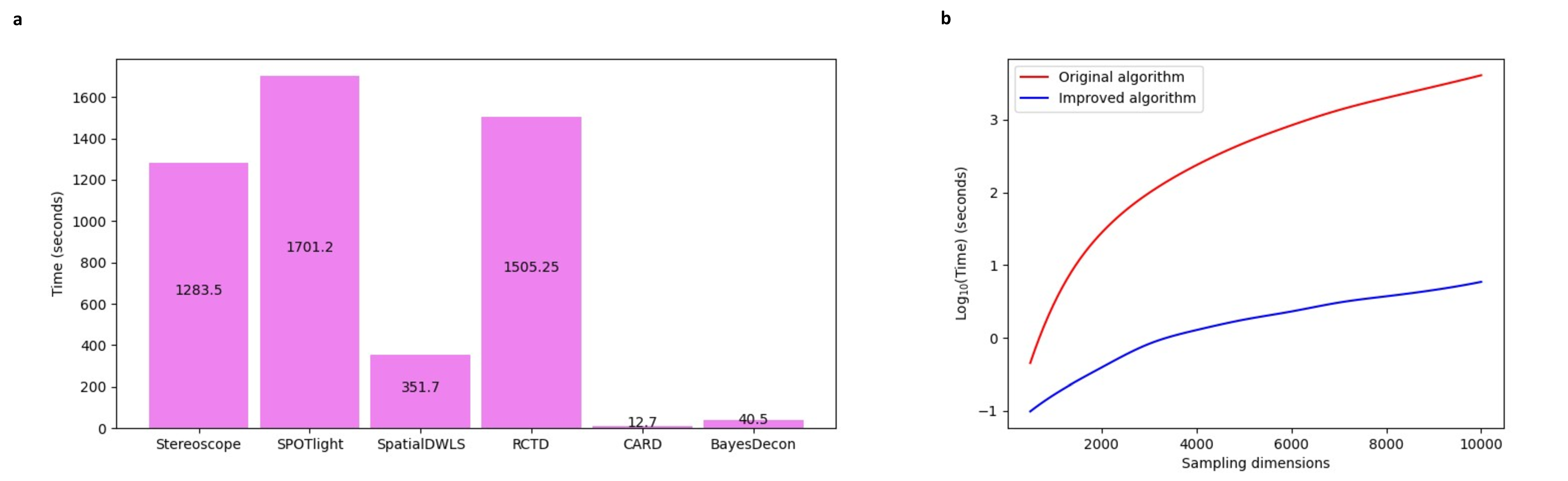}
\caption{Computational efficiency. \textbf{a}, We use the PDAC-A data to compare the computational efficiency of the six compared methods. All the methods are implemented with 11th Gen Intel(R) Core(TM) i7-11800H @ 2.30GHz and Stereoscope also utilizes NVIDIA GeForce RTX 3080 Laptop GPU. We run Stereoscope for 5000 epochs for both the spatial transcriptomics and scRNA-seq data, and the time of BASIN means that for one sample. The other four methods are implemented in the default settings. \textbf{b}, The times of generating one sample from a truncated multivariate normal distribution vs sampling dimensions. Red: the algorithm in \cite{li2015efficient}. Blue: our improved algorithm. In practice we generate one sample by ignoring first five samples and take the sixth sample.}\label{time}
\end{figure}

\begin{figure}[ht]
\centering
\includegraphics[width=1\textwidth]{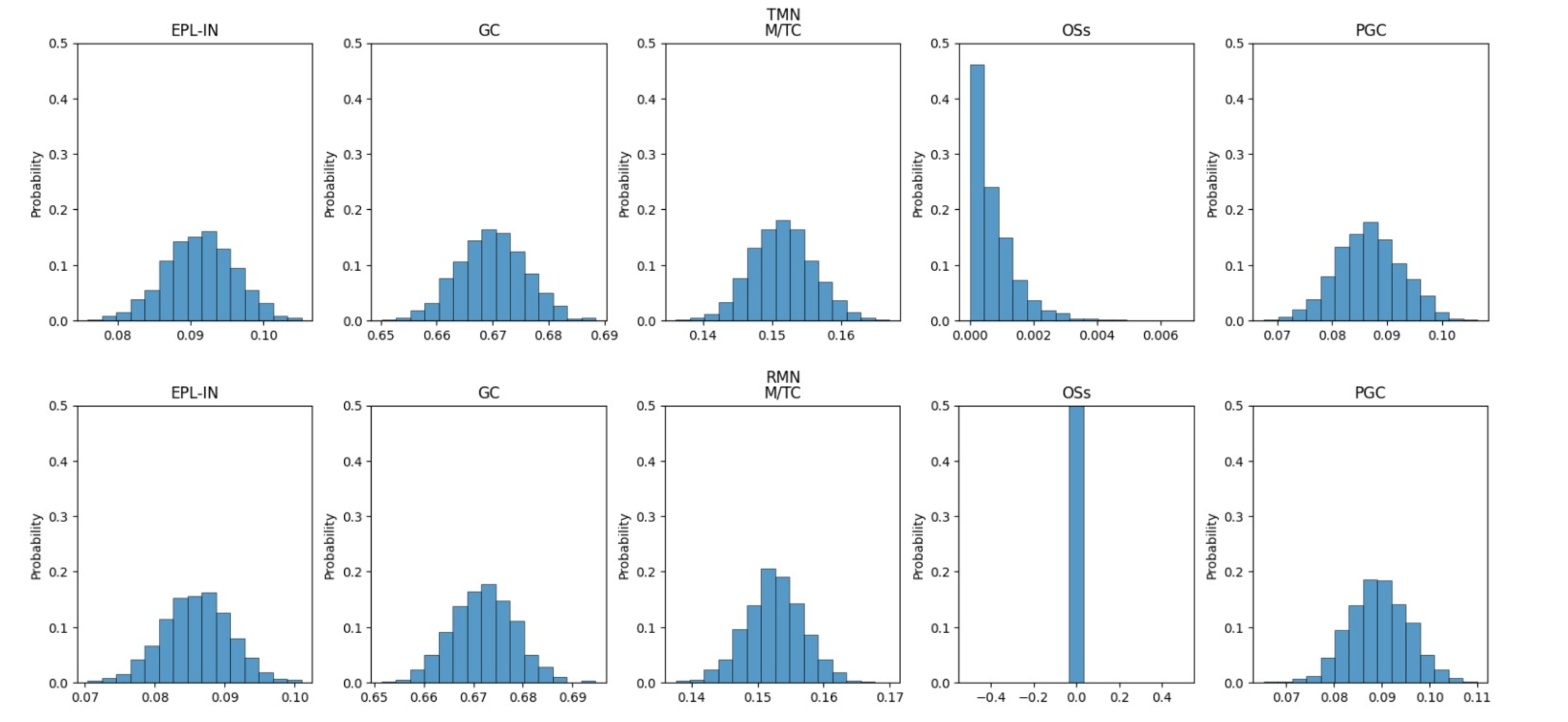}
\caption{We make an experiment to compare rectified matrix normal (RMN) and truncated matrix normal (TMN) distribution. We use the MOB data and plot the probability histograms of the cell type proportions at the same spot as in the main text, still calculated with 2000 samples. We can observe that when the mean values are far from zero (like GC and M/TC), the samples of RMN are really close to those of TMN. When the mean values are close to zero (like EPL-IN, OSs and PGC), the difference between RMN and TMN is getting obvious. However, sampling from RMN is much faster than TMN.}\label{rmn}
\end{figure}

\newpage
~\\
\newpage

\section{Details of the method}
\subsection{Matrix normal distribution}
Matrix normal distribution is a generalization of multivariate normal distribution. Suppose there is a random matrix $\mathbf{X}\in \mathbb{R}^{n\times p}$ whose entries follow normal distributions, and the mean values of $\mathbf{X}$ is a matrix $\mathbf{M}$ of size $n\times p$. Each column of $\mathbf{X}$ follows a multivariate normal distribution with covariance matrix $\mathbf{U}$ of size $n\times n$. Each row of $\mathbf{X}$ follows a multivariate normal distribution with covariance matrix $\mathbf{V}$ of size $p\times p$. Then $\mathbf{X}$ follows a matrix normal distribution $\mathbf{X} \sim \mathcal{MN}(\mathbf{M}, \mathbf{U}, \mathbf{V})$ where $\mathbf{U}$ and $\mathbf{V}$ are column and row covariances. Its probability density function is 
\begin{flalign}
P(\mathbf{X}) = \frac{\exp(-\frac{1}{2} \textbf{tr}(\mathbf{V}^{-1} (\mathbf{X}-\mathbf{M})^T  \mathbf{U}^{-1} (\mathbf{X}-\mathbf{M}) )}
{(2\pi)^{np/2} \det(\mathbf{U})^{p/2} \det(\mathbf{V})^{n/2}}
\end{flalign}
On the other hand, if the probability density function of a random matrix $\mathbf{X}\in \mathbb{R}^{n\times p}$ satisfies 
\begin{align}
 P(\mathbf{X}) \propto \exp(-\frac{1}{2}\textbf{tr}(\mathbf{AX}^T \mathbf{BX - 2CX}))  
\end{align}
where $\mathbf{A}$ and $\mathbf{B}$ are symmetric, then $\mathbf{X}$ can be represented in form of matrix normal distribution:
\begin{align*}
P(\mathbf{X}) &\propto \exp(-\frac{1}{2}
\textbf{tr}(\mathbf{AX}^T \mathbf{BX - 2CX})) \\
& \propto \exp(-\frac{1}{2}
\textbf{tr}(\mathbf{AX}^T \mathbf{BX - 2AA}^{-1}\mathbf{CB}^{-1}\mathbf{BX}))\\
& \propto \exp(-\frac{1}{2}
\textbf{tr}(\mathbf{AX}^T \mathbf{BX - 2AA}^{-1}\mathbf{CB}^{-1}\mathbf{BX}
+ \mathbf{A(A}^{-1}\mathbf{CB}^{-1}) 
\mathbf{B(B}^{-T} \mathbf{C}^T \mathbf{A}^{-T}) )) \\
& \propto \exp(-\frac{1}{2}
\textbf{tr}(\mathbf{A(X} -\mathbf{(B}^{-T} \mathbf{C}^T \mathbf{A}^{-T} )^T
\mathbf{B} \mathbf{(X} -\mathbf{(B}^{-T} \mathbf{C}^T \mathbf{A}^{-T} ) )
\end{align*}
which is equivalent to the pdf of the matrix normal distribution
\begin{align}
\mathcal{MN}(\mathbf{B}^{-T} \mathbf{C}^T \mathbf{A}^{-T}, \mathbf{B}^{-1}, \mathbf{A}^{-1})
\label{prop}
\end{align}

\subsection{Posterior of $\mathbf{V}$}
In the main text we have defined the following likelihood and priors
\begin{align}
& \mathbf{X} | \mathbf{V}, \sigma^2 \sim \mathcal{MN}(\mathbf{BV}, \sigma^2 \mathbf{I_n}, \mathbf{L}^{-1}) 
\end{align}
\begin{align}
& \mathbf{V}_{ij} | \eta \sim \text{Exp}(\eta) 
\qquad 1\le i\le c \quad 1\le j\le p
\end{align}
Their probability density function (pdf) of likelihood satisfies
\begin{align}
\begin{split}
P(\mathbf{X} | \mathbf{V}, \sigma^2) & \propto 
\frac{1}{|\sigma^2|^{np/2} }
\exp( -\frac{1}{2\sigma^2} \textbf{tr}(\mathbf{L(X-BV)}^T\mathbf{(X-BV)))} \\
& \propto 
\exp( -\frac{1}{2\sigma^2} \textbf{tr}(\mathbf{L(X-BV)}^T\mathbf{(X-BV)))} \\
& \propto \exp( -\frac{1}{2\sigma^2} 
\textbf{tr}( \mathbf{LV}^T\mathbf{B}^T \mathbf{BV - 2LX}^T \mathbf{BV + LX}^T\mathbf{X}   )) \\
& \propto \exp( -\frac{1}{2\sigma^2} 
\textbf{tr}( \mathbf{LV}^T\mathbf{B}^T \mathbf{BV - 2LX}^T \mathbf{BV}))
\end{split} 
\label{likelihood}
\end{align}
The pdf of $\mathbf{V}_{ij} | \eta$ satisfies 
\begin{align}
& P(\mathbf{V}_{ij} | \eta) = 
\eta \exp(-\eta \mathbf{V}_{ij}) \quad \mathbf{V}_{ij}\ge 0
\end{align}
\begin{align}
P(\mathbf{V} | \eta) = \prod_{i,j} P(\mathbf{V}_{ij} | \eta) = 
\eta^{cp} \exp(-\eta \sum \mathbf{V}_{ij}) = 
\eta^{cp} \exp(-\eta \textbf{tr}(\mathbf{J_V}^T \mathbf{V})) 
\label{prior_v2}
\end{align}
where $\mathbf{J_V}$ is an "all ones matrix" of the same size as $\mathbf{V}$. According to Bayes's formula,
\begin{align*}
\begin{split}
P(\mathbf{V} | \mathbf{X}) &\propto P(\mathbf{X} | \mathbf{V})P(\mathbf{V}) \\
& \propto \exp( -\frac{1}{2\sigma^2} 
\textbf{tr}( \mathbf{LV}^T\mathbf{B}^T \mathbf{BV - 2LX}^T \mathbf{BV}  ))
\exp(-\eta \textbf{tr}(\mathbf{J_V}^T \mathbf{V})) \\
& \propto \exp( -\frac{1}{2\sigma^2} 
\textbf{tr}( \mathbf{LV}^T\mathbf{B}^T \mathbf{BV - 2LX}^T \mathbf{BV} +
2\sigma^2 \eta\mathbf{J_V}^T \mathbf{V})) \\
& \propto \exp( -\frac{1}{2\sigma^2} 
\textbf{tr}( \mathbf{LV}^T \mathbf{B}^T \mathbf{BV - 2(LX}^T \mathbf{B} - \sigma^2 \eta\mathbf{J_V}^T) \mathbf{V})) 
\end{split}
\end{align*}
Applying the equation~\ref{prop}, it satisfies a matrix normal distribution with the mean matrix: 
\begin{align}
\begin{split}
\mathbf{M} &= (\mathbf{B}^T \mathbf{B})^{-T}
\mathbf{(LX}^T \mathbf{B} - \sigma^2 \eta\mathbf{J_V}^T)^T \mathbf{L}^{-T} \\
&= (\mathbf{B}^T \mathbf{B})^{-1}
\mathbf{(B}^T \mathbf{XL} - \sigma^2 \eta\mathbf{J_V}) \mathbf{L}^{-1} \\
&= (\mathbf{B}^T \mathbf{B})^{-1}
\mathbf{(B}^T \mathbf{X} - \sigma^2 \eta\mathbf{J_V}\mathbf{L}^{-1})
\end{split}
\end{align}
The two covariance matrices are $(\mathbf{B}^T \mathbf{B})^{-1}$ and $\mathbf{L}^{-1}$, and the variance $\sigma^2$ can be merged with either of them. Considering that the exponential distribution is constraint to be nonnegative, the posterior of $\mathbf{V}$ follows the truncated matrix normal distribution:
\begin{align}
\mathbf{V} | \mathbf{X} \sim \mathcal{TMN}
((\mathbf{B}^T\mathbf{B})^{-1}(\mathbf{B}^T\mathbf{X} - \sigma^2 \eta \mathbf{J_V} \mathbf{L}^{-1}), 
\; (\mathbf{B}^T\mathbf{B})^{-1} \sigma^2, 
\;  \mathbf{L}^{-1}) \quad (\mathbf{V}\ge 0)
\end{align}

\subsection{Posterior of $\sigma^2$}
With $\sigma^2 \sim \text{Inverse-Gamma}(a_2, b_2)$, the pdf satisfies
\begin{align}
P(\sigma^2) \propto |\sigma^2|^{-a_2-1}\exp(-\frac{b_2}{\sigma^2})
\end{align}
According to equation~\ref{likelihood} we have (let $\mathbf{E=X-BV}$)
\begin{align}
  P(\mathbf{X} | \sigma^2)  \propto 
  \frac{1}{|\sigma^2|^{np/2} }
\exp( -\frac{1}{2\sigma^2} \textbf{tr}(\mathbf{LE}^T\mathbf{E))}
\end{align}
Then
\begin{align*}
P(\sigma^2|\mathbf{X}) & \propto P(\mathbf{X}| \sigma^2) P(\sigma^2) \\
& \propto |\sigma^2|^{-np/2 - a_2 - 1} \exp(\frac{1}{\sigma^2}
(-b_2 - \frac{\textbf{tr}( \mathbf{LE}^T \mathbf{E})}{2} ))
\end{align*}
So the posterior of $\sigma^2$ follows the inverse Gamma distribution
\begin{align}
\sigma^2 | \mathbf{X} \sim \text{Inverse-Gamma}(a_2 + np/2, \quad b_2 + \frac{\textbf{tr}(\mathbf{LE}^T \mathbf{E})}{2} ))
\end{align}

\subsection{Posterior of $\eta$}
With $\eta \sim \text{Gamma}(a_3, b_3)$, the pdf satisfies
\begin{align}
P(\eta) \propto \eta^{a_3-1} \exp(- \frac{\eta}{b_3}) 
\end{align}
According to equation~\ref{prior_v2} we have
\begin{align}
P(\mathbf{V} | \eta) \propto 
\eta^{cp} \exp(-\eta \textbf{tr}(\mathbf{J_V}^T \mathbf{V})) 
\end{align}
\begin{align}
& P(\eta| \mathbf{V}) \propto P(\mathbf{V}|\eta)P(\eta) 
\propto \eta^{cp + a_3 - 1} \exp(-\eta (\textbf{tr}(\mathbf{J_V}^T \mathbf{V})+ \frac{1}{b_3})) 
\end{align}
So the posterior of $\eta$ follows the Gamma distribution
\begin{align}
\eta| \mathbf{V} \sim \text{Gamma}
(a_3 + cp, \quad 1/(\textbf{tr}(\mathbf{J_V}^T \mathbf{V})+ \frac{1}{b_3}))
\end{align}

\section{Sampling from nonnegative multivariate normal distribution}
We sample from truncated multivariate normal distribution (TMVN) based on the method proposed in \cite{li2015efficient} in which a Gibbs sampler is built. To improve the efficiency, we only consider the case that the lower bound is zero and the upper bound is infinity. Here we briefly describe the algorithm and how we improve it. Given an n-dimensional multivariate normal distribution truncated in $[0, \infty)$:
\begin{align}
 \mathbf{w} \sim \mathcal{TMVN}(\mathbf{\mu}, \mathbf{\Sigma})
 \quad \mathbf{w}\ge \mathbf{0}
\end{align}
whose covariance $\mathbf{\Sigma}$ is positive-definite, one can find the Choleskey decomposition $\mathbf{\Sigma} = \mathbf{L}\mathbf{L}^T$ and introduce the transformation $\mathbf{x} = \mathbf{L}^{-1} (\mathbf{w} - \mathbf{\mu})$. The new random variable $\mathbf{x}$ follows
\begin{align}
 \mathbf{x} \sim \mathcal{TMVN}(\mathbf{0}, \mathbf{I})
 \quad \mathbf{Lx}\ge -\mathbf{\mu}
\end{align}
Then it's proved in \cite{li2015efficient} that the ith variate of $\mathbf{x}$ follows the conditional distribution:
\begin{align}
\label{tuvn}
 \mathbf{x_i} | \mathbf{x_{-i}} \sim \mathcal{TN}(0, 1)
 \quad \mathbf{L_i x_i}\ge -\mathbf{\mu} - \mathbf{L_{-i} x_{-i}}
\end{align}
where $\mathbf{x_{-i}} = (\mathbf{x_1}, \cdots,  \mathbf{x_{i-1}},  \mathbf{x_{i+1}}, \cdots,  \mathbf{x_n}) $ represents the $(n-1)\times 1$ vector by removing the ith entry of $\mathbf{x}$, and $\mathbf{L_{-i}}$ represents the $n\times (n-1)$ matrix by removing the ith column of $\mathbf{L}$. Sampling from the truncated univariate normal distribution in Equation~\ref{tuvn} is not discussed here. The original algorithm can be summerized as Algorithm~\ref{original}:
\begin{algorithm}
\caption{}\label{original}
\begin{algorithmic}
\State \textbf{Given:} $\mathbf{\mu}$, $\mathbf{\Sigma}$, sample size $T$, burn-in period $T_b$ 
\State \textbf{Initialize:} choose $\mathbf{x}^0$, compute $\mathbf{\Sigma} = \mathbf{L}\mathbf{L}^T$
\For{$t$ = 1 to $T$}
\For{$i$ = 1 to $n$}
    \State Define $ \mathbf{x^t_{-i}} = (\mathbf{x^t_1}, \cdots,  \mathbf{x^t_{i-1}},  \mathbf{x^{t-1}_{i+1}}, \cdots,  \mathbf{x^{t-1}_n}) $
    \State Sample from $\mathbf{x^t_i} | \mathbf{x^t_{-i}} \sim \mathcal{TN}(0, 1) \quad \mathbf{L_i x^t_i}\ge -\mathbf{\mu} - \mathbf{L_{-i} x^t_{-i}}$
    \State Update $ \mathbf{x^t} = (\mathbf{x^t_1}, \cdots,  \mathbf{x^t_{i}},  \mathbf{x^{t-1}_{i+1}}, \cdots,  \mathbf{x^{t-1}_n}) $
\EndFor
\State $\mathbf{w^t} = \mathbf{L}\mathbf{x^t} + \mathbf{\mu} $
\EndFor
\State Discard $\mathbf{w}^1 \cdots \mathbf{w}^{T_b}$
\end{algorithmic}
\end{algorithm}

We improve the algorithm by replacing the matrix-vector multiplication in each step with scalar-vector multiplication. Therefore the larger the sampling dimension is, the more computation time we can save (Fig.~\ref{time}b). The improved sampling algorithm is summarized in Algorithm~\ref{improved}:
\begin{algorithm}
\caption{}\label{improved}
\begin{algorithmic}
\State \textbf{Given:} $\mathbf{\mu}$, $\mathbf{\Sigma}$, sample size $T$, burn-in period $T_b$ 
\State \textbf{Initialize:} choose $\mathbf{x}^0$, compute $\mathbf{\Sigma} = \mathbf{L}\mathbf{L}^T$, $\mathbf{z} = \mathbf{L}\mathbf{x^0}$
\For{$t$ = 1 to $T$}
\For{$i$ = 1 to $n$}
    \State Define $ \mathbf{x^t_{-i}} = (\mathbf{x^t_1}, \cdots,  \mathbf{x^t_{i-1}},  \mathbf{x^{t-1}_{i+1}}, \cdots,  \mathbf{x^{t-1}_n}) $
    \State Sample from $\mathbf{x^t_i} | \mathbf{x^t_{-i}} \sim \mathcal{TN}(0, 1) \quad \mathbf{L_i x^t_i}\ge -\mathbf{\mu} - \mathbf{z} + \mathbf{L_i}\mathbf{x^{t-1}_i}$
    \State Update $ \mathbf{x^t} = (\mathbf{x^t_1}, \cdots,  \mathbf{x^t_{i}},  \mathbf{x^{t-1}_{i+1}}, \cdots,  \mathbf{x^{t-1}_n}) $
    \State Update $\mathbf{z} = \mathbf{z} + \mathbf{L_i}(\mathbf{x^t_i} - \mathbf{x^{t-1}_i})$
\EndFor
\State $\mathbf{w^t} = \mathbf{L}\mathbf{x^t} + \mathbf{\mu} $
\EndFor
\State Discard $\mathbf{x}^0 \cdots \mathbf{x}^{T_{b-1}}$
\end{algorithmic}
\end{algorithm}

\newpage
~\\

\section{Details of datasets}
\begin{table}[h]
\centering
\begin{tabularx}{1\textwidth}{ 
  | >{\hsize=0.4\hsize}X 
  | >{\hsize=0.1\hsize}X 
  | >{\hsize=0.1\hsize}X
  | >{\hsize=0.1\hsize}X
  | >{\hsize=0.3\hsize}X| }
\hline
 Dataset & \# genes & \# spots & H\&E image & scRNA-seq data \\
 \hline
 MOB (Replicate 12) \cite{staahl2016visualization} & 16034 & 282 & Yes & GSE121891 \cite{tepe2018single} \\ \hline
 Human PDAC-A-1 (GSM3036911) \cite{moncada2020integrating}& 19738 & 428 & Yes & GSE111672 (PDAC-A) \cite{moncada2020integrating} \\ \hline
 Human PDAC-B (GSM3405534) \cite{moncada2020integrating}& 19738 & 224 & No & GSE111672 (PDAC-B) \cite{moncada2020integrating} \\ \hline
 Mouse brain cortex (seqFISH+) \cite{eng2019transcriptome} & 10000 & 524 & No & GSE102827 \cite{hrvatin2018single} \\ \hline
\end{tabularx}
\caption{Spatial transcriptomics data we used in our studies.}
\end{table}

\begin{table}[h]
\centering
\begin{tabularx}{0.75\textwidth}
{|>{\hsize=0.8\hsize}X|>{\hsize=0.2\hsize}X|} \hline
 Cell type & \#  \\
 \hline
 All & 12801 \\ \hline
 EPL-IN (external plexiform layer interneurons) & 161 \\ \hline
  GC (granule cells)  & 8614 \\ \hline
MT-C (mitral and tufted cells) & 1133 \\ \hline
OSNs (olfactory sensory neurons) & 1200 \\ \hline
   PGC (periglomerular cells) & 1693 \\ \hline
\end{tabularx}
\caption{Detailed cell type information of the scRNA-seq dataset GSE121891.}
\end{table}

\begin{table}[h]
\centering
\begin{tabularx}{0.75\textwidth}
{|>{\hsize=0.8\hsize}X|>{\hsize=0.2\hsize}X|} \hline
 Cell type & \#  \\
 \hline
 All & 48266 \\ \hline
 Astrocytes 	&7039 \\ \hline
Endothelial cells	&4071 \\ \hline
Excitatory cells layer 2\&3	&2963 \\ \hline
Excitatory cells layer 4	&3198 \\ \hline
Excitatory cells layer 5	&3793 \\ \hline
Excitatory cells layer 6	&3276 \\ \hline
Excitatory cells	&1057 \\ \hline
Interneurons	&936 \\ \hline
Macrophage	&537 \\ \hline
Microglia	&10158 \\ \hline
Mural	&782 \\ \hline
Oligodendrocytes&	10456 \\ \hline
\end{tabularx}
\caption{Detailed cell type information of the scRNA-seq dataset GSE102827.}
\end{table}

\begin{table}[h]
\centering
\begin{tabularx}{0.75\textwidth}
{|>{\hsize=0.8\hsize}X|>{\hsize=0.2\hsize}X|} \hline
 Cell type & \#  \\
 \hline
 All & 1926 \\ \hline
Acinar cells &	13 \\ \hline
Cancer clone A&	126\\ \hline
Cancer clone B&	170\\ \hline
Ductal - APOL high/hypoxic &	215\\ \hline
Ductal - CRISP3 high/centroacinar like &	529\\ \hline
Ductal - MHC Class II &	287\\ \hline
Ductal - terminal ductal like &	350\\ \hline
Endocrine cells &	3\\ \hline
Endothelial cells &	11\\ \hline
Fibroblasts &	5\\ \hline
Macrophages A&	21\\ \hline
Macrophages B&	19\\ \hline
Mast cells &	14\\ \hline
mDCs A&	12\\ \hline
mDCs B&	33\\ \hline
Monocytes &	18\\ \hline
pDCs&	13\\ \hline
RBCs&	15\\ \hline
T cells \& NK cells  &	40\\ \hline
Tuft cells	&32\\ \hline
\end{tabularx}
\caption{Detailed cell type information of the scRNA-seq dataset GSE111672 (PDAC-A).}
\end{table}

\begin{table}[h]
\centering
\begin{tabularx}{0.75\textwidth}
{|>{\hsize=0.8\hsize}X|>{\hsize=0.2\hsize}X|} \hline
 Cell type & \#  \\
 \hline
 All & 1733 \\ \hline
Acinar cells &	6 \\ \hline
Cancer clone A&	339\\ \hline
Ductal - CRISP3 high/centroacinar like &	152\\ \hline
Ductal - MHC Class II &	211\\ \hline
Ductal - terminal ductal like &	736\\ \hline
Endocrine cells &	13\\ \hline
Endothelial cells &	159\\ \hline
Macrophages &	9\\ \hline
Mast cells &	13\\ \hline
mDCs &	35\\ \hline
Monocytes &	20\\ \hline
RBCs&	3\\ \hline
Tuft cells	&37\\ \hline
\end{tabularx}
\caption{Detailed cell type information of the scRNA-seq dataset GSE111672 (PDAC-B).}
\end{table}

\end{document}